\documentclass[12pt,onecolumn]{article}

\usepackage[T1]{fontenc}
\usepackage[utf8]{inputenc}

\usepackage{xr}
\usepackage[margin=1in]{geometry}
\usepackage{setspace}
\usepackage{amsmath,amssymb,amsthm}
\usepackage{graphicx}
\usepackage{booktabs}
\usepackage{enumitem}
\usepackage[authoryear,round]{natbib}
\setcitestyle{aysep={}}

\newtheorem{theorem}{Theorem}
\newtheorem{proposition}{Proposition}
\newtheorem{lemma}{Lemma}
\newtheorem{corollary}{Corollary}
\theoremstyle{remark}

\newcommand{\Qh}{\widehat Q}
\newcommand{\Sh}{\widehat S}
\newcommand{\ch}{\widehat c}
\newcommand{\dhat}{\widehat d}
\newcommand{\enc}{\operatorname{enc}}
\newcommand{\supp}{\operatorname{supp}}
\newcommand{\rint}{\operatorname{rint}}

\title{Exact Quantum Search for Standard Quadratic Programming: Explicit Reversible Oracles and Resource Counts}
\author{Francesco Cesarone$^{1}$, Massimiliano Corradini$^{1}$, Andrea Scozzari$^{2,*}$\\[1ex]
{\small $^1$Roma Tre University -- Department of Business Studies}\\
{\small Via Silvio D'Amico, 77 -- 00145, Rome, Italy}\\
{\small francesco.cesarone@uniroma3.it, massimiliano.corradini@uniroma3.it}\\[1ex]
{\small $^2$Dept.\ of Economic, Psychological, Communication, Educational, and Motor Sciences}\\
{\small University Niccol\`o Cusano, Via Don Carlo Gnocchi n.\ 3 -- Rome, Italy}\\
{\small andrea.scozzari@unicusano.it}\\[1ex]
{\small $^*$Corresponding author. ORCID 0000-0003-3038-3957}}
\date{}

\begin{document}
\maketitle

\begin{abstract}
Standard quadratic programming (StQP) minimizes a possibly nonconvex quadratic form over the standard simplex and is NP-hard. Despite its simple formulation, it encompasses a broad range of continuous quadratic optimization problems, including quadratic optimization over polytopes with explicitly available vertices. We develop a gate-based quantum approach to StQP that requires no discretization of the decision variables and preserves the continuous optimal value. A known structural result reduces StQP to a finite search over admissible supports, to which quantum minimum finding applies, provided that admissibility and objective comparison can be evaluated coherently. To the best of our knowledge, we provide the first explicit reversible oracles, with fully accounted costs, for a continuous nonconvex quadratic optimization problem. For rational input, these predicates reduce to sign tests on integers of polynomially bounded bit length, implemented by clean reversible circuits based on fraction-free elimination. The construction returns an exact rational global minimizer with prescribed success probability, with explicit gate, qubit and classical-processing bounds, also over a finite gate set. The resource bounds are polynomial for a fixed cardinality bound $K$ (for instance, minimum-variance portfolio selection with a fixed maximum number of assets) or a fixed bound on the clique number of the associated convexity graph, with a quadratic reduction in candidate queries within each layer. The predicates and 102 generated circuits are verified computationally. The proposed construction serves as an exact, fully resource-accounted baseline.
\end{abstract}


\maketitle

\section{Introduction}
\label{sec:intro}

Standard quadratic programming (StQP) asks for the minimum of a possibly
nonconvex quadratic form over the standard simplex,
\begin{equation}\label{eq:stqp}
(\mathrm{StQP})\qquad \min_{x\in\Delta^n} x^TQx,
\end{equation}
where $Q$ is a symmetric $n\times n$ matrix and
$\Delta^n=\{x\in\mathbb R^n_+ : e^Tx=1\}$ is the standard simplex, with $e$ the
all-ones vector. The problem is NP-hard, and even detecting local solutions is
NP-hard \citep{Bomze1998}. It contains the maximum-clique problem as a special
case and arises in biology, economics and portfolio optimization
\citep[see, e.g.,][]{Bomze2022}. Exact and heuristic classical methods are surveyed in
\citet{Locatelli2021}; see also \citet{Bisori2022,Gondzio2021,Liuzzi2019,ScozzariTardella2008}.

\noindent The apparent simplicity of this formulation should not be mistaken for a narrow modeling scope. Linear and constant terms can be absorbed into the quadratic form by using the simplex identity $e^T x=1$, since $x^TQx+2q^Tx=x^T(Q+eq^T+qe^T)x$ on $\Delta^n$
\citep{Gondzio2021}. More generally, minimizing a quadratic function over a polytope with $m$ explicitly available vertices admits an exact reformulation
as an StQP in $m$ variables, by expressing each feasible point as a convex combination of these vertices \citep{ScozzariTardella2008}. For example, quadratic optimization over the $\ell_1$-ball in $\mathbb{R}^n$ reduces to an StQP in $2n$ variables. Thus, StQP provides a common formulation for a broad range of
continuous, possibly nonconvex quadratic optimization problems. The size of the reformulation must nevertheless be taken into account: a polytope specified by linear inequalities may have exponentially many vertices.

This modeling scope motivates the approach developed here. We construct a gate-based quantum algorithm that preserves the continuous optimization problem, rather than replacing its variables by a finite grid. For rational input, the algorithm returns an exact rational global minimizer with prescribed success probability, using explicitly constructed reversible admissibility and objective-comparison oracles with fully accounted resource costs.

\paragraph{Quantum approaches to optimization.}
Quantum optimization comprises several paradigms. Adiabatic computation and
annealing encode the objective in a Hamiltonian evolution
\citep{Albash2018}; \citet{Bochkarev2026} review quantum approaches to
discrete optimization and their hardware limitations. Variational methods
such as QAOA \citep{Farhi2014} have guarantees only for specific problem
classes and parameter regimes \citep{Hadfield2023,Wang2018}. Both families
typically require the problem to be cast as a binary polynomial, for instance by discretizing continuous variables \citep{Iftakher2023,Yuan2024}, which need not preserve the continuous optimal value. A different
route is search-based: the minimum-finding algorithm of
\citet{DurrHoyer1996} and Grover adaptive search \citep{Baritompa2005,
Gilliam2021} find the minimum of a function on a finite domain with a
quadratic reduction in the number of function evaluations. The usefulness of such algorithms ultimately depends on the evaluation oracle they are provided with; however, in most optimization applications, this oracle is assumed to be available at unit cost. This is the point at which our paper starts.

\paragraph{StQP as a search problem.}
StQP has a finite combinatorial structure that fits search-based algorithms
without discretization. By the results of \citet{ScozzariTardella2008} and
\citet{Tardella2011}, used in \citet{Cesarone2013}, at least one global
minimizer is supported on a face where the objective is strictly convex and
whose face minimizer lies in the relative interior. StQP is therefore a finite
minimization over \emph{admissible supports}, each given by a binary incidence
string and each determining a candidate minimizer and its value through a
linear system. This representation preserves the continuous optimal value,
and, unlike a discretized QUBO, the resulting objective is not a quadratic
function of the incidence bits. Admissible supports are cliques of the
convexity graph $G$ of $Q$, so a bound on the clique number of $G$ bounds the
cardinalities that must be searched. The same structure is used by classical
exact methods, notably the clique algorithm of \citet{ScozzariTardella2008}.
Searching over supports is therefore the quantum counterpart of an established
classical paradigm rather than an unrelated enumeration.

\paragraph{The gap: coherent predicates.}
Applying quantum minimum finding to this representation requires, for a
support in superposition, a coherent test of whether it is admissible
(positive definiteness of a reduced matrix and strict interiority of its
minimizer) and whether it improves an incumbent (an exact comparison of
rational objective values). Fixed-cardinality subsets can be indexed
efficiently \citep{Marsh2020}, but this does not index the
matrix-dependent subfamily of admissible supports. Specifying such predicates
as oracle interfaces and counting calls leaves open whether, and at what cost,
they can be implemented. Existing quantum work on related optimization
problems \citep{Nannicini2024,Yuan2026} does not settle this question for
StQP, and our approach does not rely on quantum linear-system algorithms
\citep[cf.][]{Nannicini2024}. This paper answers the question
constructively for rational input.

\paragraph{Contributions.}
We give an exact, resource-accounted implementation of quantum support
search for StQP, consisting of the following components.
\begin{enumerate}[leftmargin=*]
\item \emph{Integer certificates} (Section~\ref{sec:integer}).
Admissibility and objective comparison reduce to sign tests on integers
obtained from leading principal minors and Cramer numerators, with explicit
polynomial bounds on their bit length. We show that all leading minors are
needed: testing only the full determinant is not sufficient to certify
positive definiteness.
\item \emph{Clean reversible oracles} (Section~\ref{sec:reversible}).
We specialize classical fraction-free elimination \citep{Bareiss1968} and
back substitution into a single-pass evaluator that computes all support
certificates jointly: pivot-free elimination yields the leading principal
minors, and its first nonpositive pivot certifies inadmissibility;
fraction-free back substitution then recovers the Cramer numerators.
We embed this evaluator into clean reversible circuits with explicit gate
and workspace bounds. The circuits are total on the incidence register,
handle singular inputs, boundary coordinates and ties exactly, and access
the instance data coherently by table lookup \citep{Babbush2018}, with no
assumption of unit-cost quantum memory. A total order on each cardinality
layer, in which admissible supports precede inadmissible ones, lets
minimum finding start from an arbitrary, possibly inadmissible, incumbent
without requiring direct indexing of the admissible-support subfamily.
\item \emph{Search with explicit bounds} (Section~\ref{sec:search}). Layerwise
minimum finding over fixed-cardinality supports, with Dicke-state
preparation \citep{Bartschi2019}, returns an exact rational global minimizer
with prescribed success probability. We derive gate, qubit and
classical-processing bounds, both with exact rotations and with controlled
synthesis error over a finite gate set.
\item \emph{Verification} (Section~\ref{sec:experiments}). The integer
predicates agree with an independent rational implementation on 660
instances, and exhaustive basis-state simulation verifies the correctness and
cleanliness of 102 generated oracle circuits. For the reported resource
sweeps, exact arithmetic dominates the Toffoli count, identifying a primary
target for implementation improvements.
\end{enumerate}

\paragraph{\textbf{Scope and positioning}.}
The resource analysis applies both to unrestricted StQP and to its
cardinality-constrained variant, for any cardinality bound $K$. If $U$ is a
valid upper bound on the clique number of the convexity graph $G$, the search
is restricted to support layers $k\le U$ in the unrestricted case and
$k\le\min\{K,U\}$ in the cardinality-constrained case. This identifies two
structured regimes in which the search has polynomial resource bounds in $n$
and in the input bit length: families admitting a constant bound $U$, and
problems with a fixed cardinality bound $K$. The latter include
minimum-variance portfolio selection with a fixed maximum number of assets,
for which the method jointly determines the selected assets and their
continuous allocations without discretizing portfolio weights.

Within each cardinality layer, quantum minimum finding reduces
candidate-query complexity quadratically relative to exhaustive enumeration
of the same supports. For fixed $K\ge2$, the candidate-query bound over the
searched layers decreases from $O(n^K)$ to $O(n^{K/2}\log(1/\delta))$, where
$\delta$ is the prescribed failure probability; analogous bounds hold with
$U$ in place of $K$ when $U\ge2$ is fixed. Since exhaustive enumeration is
also polynomial in these regimes, the quantum contribution is a reduction in
candidate-query complexity, not polynomial-time solvability itself. Nor does
this reduction alone establish an end-to-end runtime advantage over
classical exact methods, which can prune the search
\citep{Bisori2022,Liuzzi2019,ScozzariTardella2008}, since each coherent query has a
nontrivial gate cost. However, especially in the non-convex case, classical exact and heuristic methods are typically evaluated through their performance on increasingly large benchmark instances and may, in some cases, fail to certify global optimality. By contrast, our construction provides, for any prescribed cardinality bound, an explicit worst-case resource estimate that depends solely on the bound itself and not on the structure of the instance, although its implementation remains beyond the capabilities of current quantum hardware at practically relevant scales.

To the best of our knowledge, this is the first work to construct explicit
reversible oracles, with fully accounted gate, qubit, and
classical-processing costs, for a \emph{continuous, nonconvex} quadratic
optimization problem. Existing search-based treatments of optimization
problems often assume the evaluation oracle, or address problems that are
binary from the outset \cite{Yuan2024,Yuan2026}. Here the oracles are constructed, proved correct,
and checked by exhaustive basis-state simulation on generated circuits.
Combined with the preservation of the continuous optimal value and the exact
recovery of a rational minimizer, they provide a resource-accounted baseline
for assessing quantum support search for this class of problems.

\vskip 7 pt
\noindent\textbf{Organization.} Section~\ref{sec:stqp} recalls the support
characterization and the convexity graph. Section~\ref{sec:integer} derives
the integer certificates, Section~\ref{sec:reversible} builds the reversible
oracles, and Section~\ref{sec:search} gives the search algorithm and its
resource bounds. Section~\ref{sec:experiments} reports the computational
validation, and Section~\ref{sec:conclusions} concludes. The appendices
contain the remaining proofs (Appendix~\ref{app:proofs}), the general
pivoted fraction-free determinant procedure used as an independent reference
(Appendix~\ref{app:bareiss-general}), further reversible-implementation
details (Appendix~\ref{app:reversible}), the synthesis-error accounting
(Appendix~\ref{app:synthesis}), and the detailed computational validation
(Appendix~\ref{app:experiments}).

\section{Support characterization of StQP}
\label{sec:stqp}
In this section, we review some well-known results for the StQP that will be used to construct the predicates underlying our quantum approach.
\subsection{Admissible supports}
\label{subsec:admissible}

We recall the structural results of \citet{Cesarone2013}, \citet{ScozzariTardella2008}, and \citet{Tardella2011}
for the StQP under an upper bound $K$ on the support cardinality:
\begin{equation}\label{eq:stqpK}
(\mathrm{StQP}_K)\qquad \min_{x\in\Delta^n}\ x^TQx
\quad\text{subject to}\quad |\supp(x)|\le K,
\end{equation}
where $K\in\{1,\dots,n\}$ and $\supp(x)=\{j: x_j>0\}$. For a nonempty
$I\subseteq\{1,\dots,n\}$, let
$\Delta^n_I=\{x\in\Delta^n: x_j=0 \text{ for } j\notin I\}$, a face of
dimension $|I|-1$ with relative interior
$\rint(\Delta^n_I)=\{x\in\Delta^n_I: x_i>0 \text{ for } i\in I\}$.
Let $Q_I$ be the principal submatrix indexed by $I$. With $k=|I|$,
minimization over $\Delta^n_I$ is equivalent to $\min_{u\in\Delta^k}u^TQ_Iu$.

\begin{proposition}[{Corollary 4 of \citet{Cesarone2013}}]\label{prop:structural}
For every $K\in\{1,\dots,n\}$, problem \eqref{eq:stqpK} admits a global
minimizer $x^*$ such that, $I=\supp(x^*)$, $|I|\le K$,
$x^*\in\rint(\Delta^n_I)$, and the quadratic form is strictly convex on
$\Delta^n_I$.
\end{proposition}

For every nonempty $I$ define
\begin{equation}\label{eq:weight}
w(I)=\min\{x^TQx: x\in\Delta^n_I\}=\min_{u\in\Delta^{|I|}}u^TQ_Iu .
\end{equation}
Let $I=\{i_1<\dots<i_k\}$ with $k\ge 2$, and partition
\[
Q_I=\begin{pmatrix}P_I & b_I\\ b_I^T & d_I\end{pmatrix},
\qquad P_I\in\mathbb R^{(k-1)\times(k-1)},\ b_I\in\mathbb R^{k-1},\ d_I\in\mathbb R .
\]
Eliminating the last coordinate through $u=(y,\,1-e_{k-1}^Ty)$ gives
$u^TQ_Iu=y^TS_Iy-2c_I^Ty+d_I$, where
\begin{equation}\label{eq:SI}
S_I=P_I-b_Ie_{k-1}^T-e_{k-1}b_I^T+d_Ie_{k-1}e_{k-1}^T,
\qquad c_I=d_Ie_{k-1}-b_I .
\end{equation}
The objective is strictly convex on $\Delta^n_I$ if and only if $S_I\succ0$.
In this case its unique minimizer on the affine hull is
\begin{equation}\label{eq:yI}
S_Iy_I^*=c_I,\qquad u_I^*=\begin{pmatrix}y_I^*\\ 1-e_{k-1}^Ty_I^*\end{pmatrix},
\end{equation}
which lies in $\rint(\Delta^k)$ if and only if
\begin{equation}\label{eq:interior}
y_I^*>0,\qquad e_{k-1}^Ty_I^*<1 .
\end{equation}
Whenever these inequalities hold, $u_I^*$ is the unique minimizer on
$\Delta^k$ and
\begin{equation}\label{eq:wI}
w(I)=d_I-c_I^Ty_I^* .
\end{equation}
For a singleton $I=\{i\}$ we set $u_I^*=1$ and $w(I)=q_{ii}$. For
$k=2,\dots,n$ define
\begin{equation}\label{eq:Ck}
\mathcal C_k=\{I: |I|=k,\ S_I\succ0,\ u_I^*>0\},
\qquad
\mathcal C_1=\{\{i\}: i=1,\dots,n\},
\qquad
\mathcal C_{\le K}=\bigcup_{k=1}^{K}\mathcal C_k ,
\end{equation}
where $u_I^*$ is defined whenever $S_I\succ0$ and vector inequalities are
componentwise. Supports in $\mathcal C_k$ are called \emph{admissible}.
Proposition~\ref{prop:structural} yields the finite reformulation
\begin{equation}\label{eq:finite}
\min_{x\in\Delta^n,\ |\supp(x)|\le K} x^TQx=\min_{I\in\mathcal C_{\le K}} w(I).
\end{equation}
A corresponding optimal point is recovered by embedding $u_I^*$ in the
coordinates of $I$. The original StQP corresponds to $K=n$. Strict convexity
on $\Delta^n_I$ does not imply that $Q_I$ is nonsingular; therefore
\eqref{eq:yI}--\eqref{eq:wI} provide the general characterization used here.

\subsection{Convexity graph and cardinality bounds}
\label{subsec:graph}

Following \citet{ScozzariTardella2008}, associate with $Q$ the graph $G=(V,E)$,
$V=\{1,\dots,n\}$, with
\begin{equation}\label{eq:graph}
E=\{\{i,j\}: i<j,\ \gamma_{ij}>0\},\qquad
\gamma_{ij}=(e_i-e_j)^TQ(e_i-e_j)=q_{ii}+q_{jj}-2q_{ij}.
\end{equation}
Strict convexity on $\Delta^n_I$ implies $\gamma_{ij}>0$ for all distinct
$i,j\in I$, so every admissible support induces a clique of $G$. If
$U\in\{1,\dots,n\}$ is any valid upper bound on the clique number
$\omega(G)$, then $\mathcal C_{\le n}=\mathcal C_{\le U}$ and
\begin{equation}\label{eq:vstar}
v^*:=\min_{x\in\Delta^n}x^TQx=\min_{I\in\mathcal C_{\le U}}w(I).
\end{equation}
The choice $U=n$ is always valid. If $G$ is empty, in particular if the
objective is concave on the simplex, then $v^*=\min_i q_{ii}$. For families of
instances whose convexity graphs have uniformly bounded clique number, $U$ can
remain independent of $n$; for instance, $U=2$ whenever $G$ is a tree with at
least two vertices. Graph sparsity alone does not guarantee a bounded clique
number, and the cost of computing a bound tighter than $U=n$ must be included
in preprocessing.

For $k\in\{1,\dots,U\}$ define the cardinality layers
\[
\mathcal S_k=\{I\subseteq\{1,\dots,n\}: |I|=k\},\qquad
N_k=|\mathcal S_k|=\binom nk,
\]
and, when $\mathcal C_k\ne\emptyset$, $v_k=\min_{I\in\mathcal C_k}w(I)$
(with $v_k=+\infty$ otherwise). Then $v^*=\min_{1\le k\le U}v_k$. Supports
are represented by $n$-bit incidence strings; the search will be performed over
the full layers $\mathcal S_k$, with admissibility decided inside the oracles,
and no list of admissible supports is constructed.

\paragraph{Application to cardinality-constrained portfolio selection.}
An important application of the cardinality-constrained
formulation \eqref{eq:stqpK} is portfolio selection \citep{Cesarone2013}.
Here, $x_i$ represents the fraction of available capital
invested in asset $i$, and the simplex constraints describe
a fully invested portfolio without short selling.
In a minimum-variance model, $Q$ is the covariance matrix
of asset returns, so that $x^TQx$ measures portfolio variance.
The constraint $|\supp(x)|\le K$ limits the number of assets
held, coupling the discrete choice of assets with the
continuous allocation of capital among them. Notably, even if $Q$ is positive definite, the presence of the cardinality constraint makes problem \eqref{eq:stqpK} non-convex. Unlike the discrete portfolio model of \citet{Yuan2024},
in which investment weights are restricted to a finite
grid and encoded by binary variables, our support-based
representation preserves continuous allocations.
It jointly determines which assets to hold and how much
capital to invest in each, subject to the cardinality bound,
without discretizing the portfolio weights.

For a cardinality bound $K\ge2$ that is constant independently of $n$,
the resulting search has polynomial resource bounds
(Corollary~\ref{cor:cardinality-search}); its quantum contribution in this
regime is the query reduction discussed in Section~\ref{subsec:comparison-baseline}.

\section{Exact integer support predicates}
\label{sec:integer}

The structural results of Section \ref{sec:stqp} hold for real $Q$. To construct the search
predicates using exact finite-register arithmetic, we henceforth assume that
$Q\in\mathbb Q^{n\times n}$ is symmetric and explicitly represented in binary. In the following sections, we integrate and adapt existing tools to construct coherent-admissibility and objective-comparison predicates that form the basis of our quantum framework for solving the StQP.

\subsection{Rational input and integer scaling}
\label{subsec:rational-input}

Write $q_{ij}=a_{ij}/b_{ij}$ with $a_{ij}\in\mathbb Z$, $b_{ij}\in\mathbb Z_{>0}$,
$1\le i\le j\le n$, and let
\[
D=\prod_{1\le i\le j\le n}b_{ij},\qquad \Qh=DQ .
\]
Then $D>0$ and $\Qh$ is an integer symmetric matrix; any smaller positive
common denominator may be used instead.

\begin{proposition}[Invariance under positive scaling]\label{prop:integer-scaling}
The matrices $Q$ and $\Qh$ define the same convexity graph and the same
admissible support families, and each admissible support yields the same
continuous minimizer for both. Moreover, $w_{\Qh}(I)=D\,w_Q(I)$ for every
nonempty $I$. Consequently, the two instances have the same global minimizers,
and all objective comparisons between admissible supports are preserved.
\end{proposition}

\begin{proof}
For $i\ne j$, $\hat q_{ii}+\hat q_{jj}-2\hat q_{ij}=D\gamma_{ij}$, so the graphs
coincide. For $|I|\ge2$, the definitions \eqref{eq:SI} give $\Sh_I=DS_I$ and
$\ch_I=Dc_I$. Thus $\Sh_I\succ0$ iff $S_I\succ0$, and then
$\Sh_I^{-1}\ch_I=S_I^{-1}c_I$, so the candidate and its interiority conditions
are unchanged. Finally, $w_{\Qh}(I)=\min_{x\in\Delta^n_I}x^T\Qh x=D\,w_Q(I)$,
and the same identity holds over $\Delta^n$. Since $D>0$, minimizers and
orderings are preserved.
\end{proof}

Let $L$ be the total binary length of the rational input. Then
$\log_2D=O(L)$ and every entry of $\Qh$ has binary length $O(L)$. We use the
entry-size parameter
\[
B=1+\Bigl\lceil\log_2\bigl(1+\max_{i,j}|\hat q_{ij}|\bigr)\Bigr\rceil,
\qquad H=2^B,
\]
so that $|\hat q_{ij}|\le H$. The cost of constructing $D$ and $\Qh$ is
classical preprocessing. Exactness throughout refers to the represented
rational instance.

\subsection{Admissibility certificates}
\label{subsec:integer-predicates}

For a support $I=\{i_1<\dots<i_k\}$, $k\ge2$, set $r=k-1$, partition $\Qh_I$
as in Section~\ref{subsec:admissible}, and let $\Sh_I,\ch_I,\dhat_I$ be the
corresponding integer data \eqref{eq:SI}. Define
\[
\delta_{I,j}=\det\bigl(\Sh_I[1\!:\!j,1\!:\!j]\bigr)\ (j=1,\dots,r),
\qquad \delta_I=\delta_{I,r},
\]
let $\Sh_I^{(\ell)}$ be obtained from $\Sh_I$ by replacing its $\ell$th column
with $\ch_I$, and set
\[
z_{I,\ell}=\det\bigl(\Sh_I^{(\ell)}\bigr)\ (\ell=1,\dots,r),
\qquad
z_{I,k}=\delta_I-\sum_{\ell=1}^{r}z_{I,\ell}.
\]
These integers are defined for every support, including supports for which
$\Sh_I$ is singular or not positive definite.

\begin{proposition}[Integer characterization of admissibility]\label{prop:integer-admissibility}
Let $|I|=k\ge2$. Then $I\in\mathcal C_k$ if and only if
\[
\delta_{I,j}>0\quad(j=1,\dots,k-1)
\qquad\text{and}\qquad
z_{I,\ell}>0\quad(\ell=1,\dots,k).
\]
In this case the face minimizer, in the ordered coordinates of $I$, is the
rational vector $u_I^*=\delta_I^{-1}(z_{I,1},\dots,z_{I,k})^T$.
\end{proposition}

\begin{proof}
By Sylvester's criterion \citep{HornJohnson2013}, $\Sh_I\succ0$ iff
$\delta_{I,j}>0$ for all $j$; then $\delta_I>0$ and $\Sh_I$ is invertible.
By Cramer's rule, $(y_I^*)_\ell=z_{I,\ell}/\delta_I$ for $\ell\le k-1$, and
$1-e_r^Ty_I^*=z_{I,k}/\delta_I$. Since $\delta_I>0$, all coordinates of $u_I^*$
are positive iff $z_{I,\ell}>0$ for all $\ell$. By
Proposition~\ref{prop:integer-scaling}, this is admissibility for $Q$.
\end{proof}

Requiring all leading minors is necessary: $\delta_I>0$ alone does not certify
positive definiteness. Sylvester's criterion is applied only to the symmetric
matrix $\Sh_I$, not to the replacement matrices. Zero minors and zero
coordinates are treated as failures, so boundary cases are decided exactly.

\subsection{Objective representation and comparison}
\label{subsec:objective}

For $|I|=k\ge2$ define
\[
p_I=\dhat_I\delta_I-\sum_{\ell=1}^{k-1}(\ch_I)_\ell\, z_{I,\ell},
\]
and for a singleton $I=\{i\}$ set $\delta_I=z_{I,1}=1$ and $p_I=\hat q_{ii}$.

\begin{proposition}[Exact objective representation and comparison]\label{prop:integer-comparison}
For every admissible $I$, $\delta_I>0$ we have
\[
w_{\Qh}(I)=\frac{p_I}{\delta_I},\qquad w_Q(I)=\frac{p_I}{D\delta_I}.
\]
Hence, for admissible $I,J$, possibly of different cardinalities,
$w_Q(I)<w_Q(J)$ iff $p_I\delta_J<p_J\delta_I$, and $w_Q(I)=w_Q(J)$ iff
$p_I\delta_J=p_J\delta_I$.
\end{proposition}

\begin{proof}
For $k\ge2$, by \eqref{eq:wI} and Proposition~\ref{prop:integer-admissibility},
$w_{\Qh}(I)=\dhat_I-\delta_I^{-1}\sum_\ell(\ch_I)_\ell z_{I,\ell}=p_I/\delta_I$.
The singleton case is immediate, and $w_Q=w_{\Qh}/D$. Since $D,\delta_I,\delta_J>0$,
the comparisons follow by cross multiplication.
\end{proof}

The pair $(p_I,\delta_I)$ need not be reduced, and $p_I$ may have any sign.
The ratio $p_I/\delta_I$ is interpreted as an objective value only for
admissible supports.

\paragraph{Size of the integer data.}
For $k\ge2$ and $r=k-1$, let
\[
t_k=r(B+2)+\Bigl\lceil\tfrac r2\log_2r\Bigr\rceil,\qquad T_k=2^{t_k},
\]
and $t_1=0$, $T_1=1$. Let $U\in\{1,\dots,n\}$ be an upper bound on the clique number
$\omega(G)$ of the convexity graph $G$, and define
\[
R_U=2(2U-1)HT_U^2,\qquad W_U=2+\lceil\log_2(1+R_U)\rceil.
\]

\begin{proposition}[Bit-length bounds]\label{prop:certificate-bit-length}
For every support $I$ with $|I|=k\ge2$, admissible or not,
\[
|\delta_{I,j}|\le T_k,\quad |z_{I,\ell}|\le T_k\ (\ell<k),\quad
|z_{I,k}|\le kT_k,\quad |p_I|\le(2k-1)HT_k .
\]
For any two nonempty supports $I,J$ with $|I|,|J|\le U$, the integer
$F(I,J)=p_I\delta_J-p_J\delta_I$ satisfies $|F(I,J)|\le R_U$. All these
integers, the partial sums defining $z_{I,k}$ and $p_I$, and the products
defining $F(I,J)$ fit in signed registers of width $W_U$, and
\[
W_U=O\bigl(U(B+\log_2(U+1))\bigr).
\]
\end{proposition}

The proof, based on Hadamard's inequality \citep{HornJohnson2013}, is given in
Appendix~\ref{app:bitlength}. The width $W_U$ bounds the scalar support data;
it is not a bound on the workspace of the determinant computations, which is
treated in Section~\ref{sec:reversible}.

\subsection{Search predicates and the layer order}
\label{subsec:search-predicates}

Fix the injective encoding $\nu(I)=\sum_{j=1}^n s_j2^{j-1}$ of a support with
incidence string $s$. For admissible $I,J$ define the tie-broken order
\[
I\prec J\iff w(I)<w(J)\ \text{or}\ \bigl(w(I)=w(J)\ \text{and}\ \nu(I)<\nu(J)\bigr).
\]
By Proposition~\ref{prop:integer-comparison}, $I\prec J$ iff
\begin{equation}\label{eq:beta}
\beta(I,J)=[F(I,J)<0]\lor\bigl([F(I,J)=0]\land[\nu(I)<\nu(J)]\bigr)=1 .
\end{equation}
For $k\ge2$, the admissibility indicator on $\mathcal S_k$ is, by
Proposition~\ref{prop:integer-admissibility},
\[
a_k(I)=\Bigl(\bigwedge_{j=1}^{k-1}[\delta_{I,j}>0]\Bigr)\land
\Bigl(\bigwedge_{\ell=1}^{k}[z_{I,\ell}>0]\Bigr),
\]
and every singleton is admissible.

Within a layer, minimum finding uses a temporary incumbent that is sampled
uniformly from $\mathcal S_k$ and may therefore be inadmissible. We use the
total order $\triangleleft_k$ on $\mathcal S_k$ in which (i) every admissible
support precedes every inadmissible one; (ii) admissible supports are ordered
by $\prec$; (iii) inadmissible supports are ordered by $\nu$. If
$\mathcal C_k\ne\emptyset$, the least element of $\triangleleft_k$ is an
admissible support attaining $v_k$. No objective value is assigned to an
inadmissible support.

For an admissible incumbent $J$ with $|J|\le U$ (possibly of a different
cardinality), the \emph{improvement predicate} on $\mathcal S_k$ is
$h_{k,J}(I)=a_k(I)\land\beta(I,J)$. For a temporary incumbent
$J\in\mathcal S_k$ with admissibility flag $\eta_J=a_k(J)$, the
\emph{layer-order predicate} is
\begin{equation}\label{eq:g}
g_{k,J}(I)=
\begin{cases}
a_k(I)\land\beta(I,J), & \eta_J=1,\\
a_k(I)\lor\bigl(\neg a_k(I)\land[\nu(I)<\nu(J)]\bigr), & \eta_J=0,
\end{cases}
\end{equation}
so that $g_{k,J}(I)=1$ iff $I\triangleleft_kJ$. Both predicates are extended
to the full incidence register by the value zero on strings of Hamming weight
different from $k$. All quantities involved are the integers of
Sections~\ref{subsec:integer-predicates}--\ref{subsec:objective}.

\section{Reversible realization of the search oracles}
\label{sec:reversible}

We now construct clean reversible circuits computing the predicates of
Section~\ref{subsec:search-predicates} on the incidence register. A circuit is
\emph{clean} if all work registers start and end in $|0\rangle$ on every
basis input; its action then extends linearly to superpositions.

The building blocks of this section are standard: fraction-free
elimination \citep{Bareiss1968}, compute--copy--uncompute embedding
\citep{Bennett1973}, and table lookup \citep{Babbush2018}. What is specific
to StQP is their combination into \emph{total} predicates on the full
cardinality layer. The circuits must decide admissibility exactly,
including singular reduced matrices, zero leading minors and zero
coordinates of the face minimizer; they must return well-defined outputs on
inadmissible supports and on strings outside the layer; and they must
compare a candidate with an incumbent that may itself be inadmissible or
belong to a different layer, with exact tie-breaking. The single-pass
evaluator below obtains all certificates required by these predicates from
one elimination, which is what makes a fixed-schedule reversible
implementation with $O(k^3)$ arithmetic operations possible.

\subsection{Fraction-free evaluation of support certificates}
\label{subsec:single-pass-eval}

The certificates of Section~\ref{sec:integer} could be obtained by
$2(k-1)$ separate determinant evaluations. Since the replacement matrices
$\Sh_I^{(\ell)}$ are not symmetric and may have zero diagonal pivots, each
such evaluation would require fraction-free elimination with row pivoting;
this general procedure, with its correctness proof and explicit bounds, is
given in Appendix~\ref{app:bareiss-general}, where it serves as the
independent determinant-based reference in our experiments. The search
predicates, however, require much less. The leading minors $\delta_{I,j}$
are those of the symmetric matrix $\Sh_I$, and the Cramer numerators
$z_{I,\ell}$ are the entries of the solution of the single system
$\Sh_Iy=\ch_I$ scaled by $\delta_I$. Since admissibility requires
$\Sh_I\succ0$, elimination without row exchanges suffices on every support
that can be admissible, and the first nonpositive pivot certifies
inadmissibility. The following proposition specializes classical
fraction-free elimination and back substitution to the support
certificates. Although the underlying procedures are standard, we state
this specialization explicitly to establish the rejection criterion, derive
bounds on intermediate quantities, and quantify the operation count required
for the reversible implementation.

\begin{proposition}[Fraction-free evaluation of support certificates]\label{prop:single-pass}
Let $k\ge2$, $r=k-1$, $\beta=B+2$, let $S=\Sh_I\in\mathbb Z^{r\times r}$ be
symmetric and $c=\ch_I\in\mathbb Z^r$, with all entries of $N=[S\,|\,c]$ of
absolute value at most $2^\beta$. Put $A\leftarrow N$, $d\leftarrow1$. For
$s=1,\dots,r$:
\begin{enumerate}[leftmargin=*]
\item if $A_{ss}\le0$, return FAIL;
\item
If $s<r$, for all $i>s$ and $j>s$ ($j\leq r+1$), set
\[
A_{ij}\leftarrow
\frac{A_{ss}A_{ij}-A_{is}A_{sj}}{d},
\]
with $A_{ss}$, $d$, row $s$, and the entries $A_{is}$
unchanged during these updates.
After completing all updates, set $A_{is}\leftarrow 0$
for all $i>s$, and then set $d\leftarrow A_{ss}$.
\end{enumerate}
Let $U_{sj}=A_{sj}$ be the frozen entries of row $s$ at stage $s$, $u_s=U_{s,r+1}$,
and, if no FAIL occurred, define $\zeta_r=u_r$ and, for $s=r-1,\dots,1$,
\[
\zeta_s=\frac{\delta_I u_s-\sum_{j=s+1}^{r}U_{sj}\zeta_j}{U_{ss}}.
\]
Then:
\begin{enumerate}[label=(\alph*),leftmargin=*]
\item the procedure returns FAIL iff $S\not\succ0$;
\item if $S\succ0$, then $U_{ss}=\delta_{I,s}$ for $s=1,\dots,r$, and every
division is exact, with positive divisor;
\item if $S\succ0$, then $\zeta_s=z_{I,s}$ for $s=1,\dots,r$;
\item every stored entry is bounded by $T_k$ in absolute value, every
numerator before division by $2T_k^2$ in the elimination and by $rT_k^2$ in
the back substitution, so signed registers of width
$w'_k=2t_k+\lceil\log_2(k+1)\rceil+2$ suffice;
\item On executions that do not return \textnormal{FAIL}, the procedure
uses exactly $(r-1)r(r+1)/3$ elimination updates and $O(r^2)$
further arithmetic operations. On every input, these quantities
are upper bounds, so the arithmetic cost is $O(k^3)$.
\end{enumerate}
\end{proposition}

The proof, based on the classical bordered-minor invariant of fraction-free
elimination \citep{Bareiss1968}, is given in Appendix~\ref{app:single-pass}.

For a support of cardinality $k\ge2$, the procedure is applied to
$[\Sh_I\,|\,\ch_I]$. The successive pivots give all leading minors
$\delta_{I,1},\dots,\delta_{I,k-1}$, with the first nonpositive pivot
certifying that $I$ is inadmissible. If all pivot tests pass, fraction-free
back substitution gives all Cramer numerators $z_{I,1},\dots,z_{I,k-1}$;
then $z_{I,k}$ and $p_I$ follow from their definitions. On executions that
pass all pivot tests, the procedure uses exactly $(k-2)(k-1)k/3$ elimination
updates and $O(k^2)$ further arithmetic operations; these quantities are
upper bounds on every input. Thus the total cost is $O(k^3)$ arithmetic
operations and $O\bigl(k^5(B+\log_2(k+1))^2\bigr)$ bit operations with
schoolbook arithmetic \citep{BrentZimmermann2010}. Computing the $2(k-1)$
determinants separately with the pivoted procedure of
Appendix~\ref{app:bareiss-general} would require $O(k^4)$ arithmetic
operations. This joint certificate evaluation is the basis of the clean
reversible support-data evaluator constructed next.

\subsection{Reversible support-data evaluation}
\label{subsec:reversible-support-data}

We now turn the integer procedure of Proposition~\ref{prop:single-pass} into a clean reversible circuit. The tool is the standard embedding of Boolean circuits into reversible ones, which follows the compute--copy--uncompute construction of \citet{Bennett1973} (see also \citealt{NielsenChuang2010}).

\begin{lemma}[Clean reversible embedding]\label{lem:clean-reversible-embedding}
Let $f:\{0,1\}^a\to\{0,1\}^b$ be computed by a bounded-fan-in Boolean circuit
of size $S$. There is a circuit over NOT, CNOT and Toffoli gates with
$|x\rangle|y\rangle|0^q\rangle\mapsto|x\rangle|y\oplus f(x)\rangle|0^q\rangle$,
using $O(S+b)$ gates and $q=O(S+b)$ clean work qubits.
\end{lemma}

The copy is a bitwise XOR in the computational basis and does not involve cloning of arbitrary states.

\noindent For $k\ge2$ and $r=k-1$, let $M=\Qh_I$ be an already loaded ordered
support matrix. Define the data
\[
\mathcal D_k(M)=\bigl(\delta_{I,1},\dots,\delta_{I,k-1},\,z_{I,1},\dots,z_{I,k},\,p_I,\,a_k(I)\bigr)
\]
as follows: if $\Sh_I\succ0$, then $\delta_{I,j}$, $z_{I,\ell}$ and $p_I$ are
the integers of Section~\ref{sec:integer} and $a_k(I)=\bigwedge_{\ell=1}^k[z_{I,\ell}>0]$;
if $\Sh_I\not\succ0$ (equivalently, the single-pass procedure fails), all
integers are set to $0$ and $a_k(I)=0$. By Proposition~\ref{prop:certificate-bit-length}
and Proposition~\ref{prop:single-pass}, all these integers fit in a signed
register of common width
\[
w_k=\max\{W_k,\;2t_k+\lceil\log_2(k+1)\rceil+2,\;B+4\}
=O\bigl(k(B+\log_2(k+1))\bigr).
\]

\begin{theorem}[Clean support-data evaluator, single pass]\label{thm:reversible-support-data}
For $k\ge2$ there is a clean reversible circuit
\[
|M\rangle|Y\rangle|0^{q_k}\rangle\ \mapsto\ |M\rangle|Y\oplus\enc(\mathcal D_k(M))\rangle|0^{q_k}\rangle
\]
whose gate count and number of work qubits are
\[
G_{\mathrm{data}}(k,B)=O\bigl(k^3w_k^2\bigr)=O\bigl(k^5(B+\log_2(k+1))^2\bigr),\qquad
q_k=O\bigl(k^5(B+\log_2(k+1))^2\bigr).
\]
Copying only the flag $a_k(I)$ gives a clean admissibility evaluator with the
same bounds. For $k=1$, the data $\delta_I=z_{I,1}=1$, $p_I=M_{11}$, $a=1$ are
written with $O(B+1)$ gates.
\end{theorem}

\begin{proof}
We describe a fixed-schedule Boolean circuit $\mathcal B_k$ computing
$\mathcal D_k$ from $M$ and apply Lemma~\ref{lem:clean-reversible-embedding}.

\emph{Input preparation.} By \eqref{eq:SI}, $(\Sh_I)_{ab}=M_{ab}-M_{a,k}-M_{b,k}+M_{kk}$
for $a,b\le r$, $(\ch_I)_a=M_{kk}-M_{a,k}$ and $\dhat_I=M_{kk}$. These
$O(k^2)$ entries are computed with $O(k^2)$ additions and subtractions of
$w_k$-bit integers, i.e.\ $O(k^2w_k)$ gates. Entries outside $[-2^{B},2^{B}]$
are mapped to the all-zero output by an initial validity test.

\emph{Elimination.} All $r$ stages of Proposition~\ref{prop:single-pass} are
present for every input. A bit $\mathrm{Act}$ is initialized to~$1$. At stage
$s$ the circuit computes $\pi_s=\mathrm{Act}\wedge[A_{ss}>0]$, sets
$\mathrm{Act}\leftarrow\pi_s$, and applies the update
$A_{ij}\leftarrow(A_{ss}A_{ij}-A_{is}A_{sj})/d$ for $i,j>s$ ($j\le r+1$) and
$d\leftarrow A_{ss}$ only under control of $\mathrm{Act}$. There are no row
exchanges. The division circuit is total, outputting zero on a zero divisor;
by Proposition~\ref{prop:single-pass}(b), on every active update the divisor
$d=\delta_{I,s-1}$ is positive, the remainder is zero and the quotient is
representable in $w_k$ bits, so this convention does not affect valid outputs.
The pivots $A_{ss}$, $s=1,\dots,r$, are stored as $\delta_{I,s}$ when
$\mathrm{Act}=1$ and as $0$ otherwise.

\emph{Back substitution and final quantities.} Under control of $\mathrm{Act}$,
the circuit computes $\zeta_r=u_r$ and $\zeta_s$ for $s=r-1,\dots,1$ by the
recursion of Proposition~\ref{prop:single-pass}, which by part~(c) gives
$\zeta_s=z_{I,s}$ with exact division by $\delta_{I,s}>0$. It then computes
$z_{I,k}=\delta_I-\sum_{\ell=1}^{r}z_{I,\ell}$, $p_I=\dhat_I\delta_I-\sum_{\ell=1}^{r}(\ch_I)_\ell z_{I,\ell}$
and $a_k(I)=\mathrm{Act}\wedge\bigwedge_{\ell=1}^{k}[z_{I,\ell}>0]$; on
$\mathrm{Act}=0$ every output is zero. By
Proposition~\ref{prop:certificate-bit-length} no overflow occurs in these sums
and products.

Equivalently, the arithmetic may be evaluated unconditionally
using the total fixed-width operations, with every output integer
masked by the final activity flag and the admissibility flag
set to zero on failure.
This is the convention used by the generated implementation
described in Appendix~\ref{app:reversible}; it computes the same output function
and has the same asymptotic resource bounds.

\emph{Correctness.} By Proposition~\ref{prop:single-pass}(a), $\mathrm{Act}$
is still~$1$ after stage $r$ iff $\Sh_I\succ0$. In that case parts~(b) and (c)
give $\delta_{I,s}$ and $z_{I,\ell}$ for $\ell\le r$, and the definitions give
$z_{I,k}$, $p_I$ and $a_k(I)$. Otherwise all outputs are zero. Hence
$\mathcal B_k$ computes $\mathcal D_k$.

\emph{Size.} There are $(k-2)(k-1)k/3$ elimination updates, each with two
multiplications, a subtraction and a division on $w_k$-bit integers, hence
$O(w_k^2)$ gates each with schoolbook arithmetic
\citep{BrentZimmermann2010}. The back substitution uses $O(k^2)$ operations of
cost $O(w_k^2)$, the final sums and products $O(kw_k^2)$, and the pivot tests
and masking $O(kw_k)$ for each of the $O(k^2)$ controlled registers. The total
size is $S=O(k^3w_k^2)$. Applying
Lemma~\ref{lem:clean-reversible-embedding} (compute, copy $\enc(\mathcal D_k(M))$
into $Y$ by $O(kw_k)$ CNOT gates, uncompute) gives
$O(S+kw_k)$ gates and $O(S+kw_k)$ clean work qubits. Substituting
$w_k=O(k(B+\log_2(k+1)))$ gives the stated bounds.
\end{proof}

The workspace bound reflects a history-preserving embedding and is
deliberately larger than the classical storage of Bareiss elimination; more space-efficient reversible schedules are possible but not needed here.

\subsection{Coherent access to support matrices}
\label{subsec:coherent-support-matrices}

We use an explicit data-access model: $\Qh$ is fixed classical data compiled
into a table-lookup circuit, in the spirit of QROM constructions
\citep{Babbush2018}; no unit-cost quantum random-access memory is assumed.
Let $\ell=\lceil\log_2(n+1)\rceil$ and $b_Q=B+2$. Extend the table by
$q_{\mathrm{tab}}(u,v)=\hat q_{uv}$ for $1\le u,v\le n$ and $0$ otherwise.
The lookup
\[
\mathcal L_{\Qh}:|u\rangle|v\rangle|y\rangle\mapsto|u\rangle|v\rangle|y\oplus\enc_{b_Q}(q_{\mathrm{tab}}(u,v))\rangle
\]
is implemented by computing, for each fixed pair $(a,b)$, the flag
$[u=a]\land[v=b]$, applying CNOT gates from this flag to the one-bits of
$\enc_{b_Q}(\hat q_{ab})$, and uncomputing the flag. Processing all pairs
gives an exact clean circuit with $O(n^2(\ell+b_Q))$ gates and $O(\ell)$
reusable workspace. Its generation is part of classical preprocessing.

\paragraph{Ordered support extraction.}
For an incidence string $s$ of Hamming weight $k$, the loader first computes
the ordered addresses $i_1(s)<\dots<i_k(s)$ of its one-bits by means of
clean prefix counts, and sets all addresses to $0$ on strings of other
weights; the construction and its cost are detailed in
Appendix~\ref{app:reversible}.

\begin{proposition}[Clean support-matrix loader]\label{prop:clean-support-matrix-loader}
For every $k\in\{1,\dots,n\}$ there is a clean reversible circuit
\[
\mathcal P_k:|s\rangle|0\rangle|0^{k^2b_Q}\rangle\mapsto|s\rangle|v_k(s)\rangle|\enc(M_k(s))\rangle,
\]
where $M_k(s)=\Qh_{\supp(s)}$ if $|s|=k$ and $M_k(s)=0$ otherwise, with gate
count $O\bigl(nk\ell^2+k^2n^2(B+\ell)\bigr)$ and clean workspace
$O(n\ell^2+k\ell)$.
\end{proposition}

\begin{proof}
Compute prefix counts, $v_k(s)$ and the ordered addresses. For each
$(a,b)\in\{1,\dots,k\}^2$ apply $\mathcal L_{\Qh}$ to $i_a(s),i_b(s)$ and the
corresponding entry register. For diagonal entries ($a=b$), first copy the
address into a clean temporary register in the computational basis, apply the
lookup, and uncompute the copy. On invalid strings all addresses are $0$ and
the extended table writes zeros. Then reverse the address extraction and the
prefix counts. The $k^2$ lookups give the stated gate count; lookup,
comparison and arithmetic workspaces are reused sequentially.
\end{proof}

\noindent Combining the support-matrix loader of Proposition \ref{prop:clean-support-matrix-loader} with the
clean support-data evaluator of Theorem \ref{thm:reversible-support-data} yields an explicit
admissibility oracle acting directly on the incidence register.
The following bounds include coherent data access and enforce
a zero output outside the prescribed cardinality layer.

\begin{theorem}[Explicit admissibility oracle]\label{thm:incidence-admissibility-oracle}
For $k\ge2$ there is a clean reversible circuit
\[
|s\rangle|b\rangle|0\rangle\mapsto|s\rangle|b\oplus a_k(s)\rangle|0\rangle,
\]
where $a_k(s)=1$ iff $|s|=k$ and $\supp(s)\in\mathcal C_k$, with gate count
\[
O\bigl(nk\ell^2+k^2n^2(B+\ell)+k^5(B+\log_2(k+1))^2\bigr)
\]
and workspace $O\bigl(n\ell^2+k\ell+k^2(B+1)+k^5(B+\log_2(k+1))^2\bigr)$.
For $k=1$ the Hamming-weight test $[|s|=1]$ suffices.
\end{theorem}

\begin{proof}
Apply $\mathcal P_k$ and the clean admissibility evaluator of
Theorem~\ref{thm:reversible-support-data}, toggle $b$ with a Toffoli gate
controlled by $v_k(s)$ and the local flag, and reverse both circuits. The
explicit control by $v_k(s)$ ensures the predicate is zero outside the layer.
\end{proof}

No list of admissible supports is constructed: the circuit contains the
instance data, not an enumerated table of support values.

\subsection{Clean comparison and phase oracles}
\label{subsec:reversible-comparison}

The incumbent is maintained classically and fixed during each oracle call.
Its data are held in dedicated registers in computational basis states: the
admissibility flag $\eta_J$, the identifier $\nu(J)$, and, when $\eta_J=1$,
$(p_J,\delta_J)$; for an inadmissible incumbent the dummy values $p_J=0$,
$\delta_J=1$ are used. These registers are preserved by every call, so
incumbent updates require only $O(n+W_U)$ bit flips and no resynthesis of the
arithmetic or lookup circuits.

With all data sign-extended to width $W_U$, the bit $\beta(I,J)$ of
\eqref{eq:beta} is computed by two schoolbook multiplications, a subtraction,
sign and zero tests, and an $n$-bit identifier comparison. By
Proposition~\ref{prop:certificate-bit-length} no overflow occurs, even when
$J$ belongs to a different layer, and by
Lemma~\ref{lem:clean-reversible-embedding} this comparison has a clean
implementation with $O(W_U^2+n)$ gates and workspace. On inadmissible
candidates the arithmetic is evaluated as a total integer function, but its
result is masked by $a_k(I)$.

Set
\[
\Gamma_k=nk\ell^2+k^2n^2(B+\ell)+k^5(B+\log_2(k+1))^2+W_U^2+n,
\]
\[
\Xi_k=n\ell^2+k\ell+k^2(B+1)+k^5(B+\log_2(k+1))^2+W_U^2+n .
\]

\noindent Combining the support-matrix loader and support-data evaluator
with exact integer comparisons yields the following clean
comparison oracles. The construction handles both admissible
and inadmissible temporary incumbents, resolves objective ties
exactly, and includes all coherent data-access costs.

\begin{theorem}[Explicit clean comparison oracles]\label{thm:explicit-comparison-oracles}
Fix $\Qh$, $U$, and $2\le k\le U$. For every admissible incumbent $J$ with
$|J|\le U$, and for every temporary incumbent $J\in\mathcal S_k$, admissible
or not, there are clean reversible oracles
\[
|s\rangle|b\rangle|0\rangle\mapsto|s\rangle|b\oplus h_{k,J}(s)\rangle|0\rangle,
\qquad
|s\rangle|b\rangle|0\rangle\mapsto|s\rangle|b\oplus g_{k,J}(s)\rangle|0\rangle,
\]
with gate count $O(\Gamma_k)$ and workspace $O(\Xi_k)$, uniformly over the
permitted incumbents.
\end{theorem}

\begin{proof}
Apply $\mathcal P_k$ and the support-data evaluator; compute $\beta(I,J)$ and
$[\nu(I)<\nu(J)]$; compute the Boolean combination \eqref{eq:g} selected by
$\eta_J$, or $a_k\land\beta$ for $h_{k,J}$, conjoined with $v_k(s)$; copy it
into $b$; and reverse the comparison, the evaluator, and the loader. The
target is not used by the inverse computations, so all work registers return
to zero. The cost is the sum of the forward and inverse costs of the
components.
\end{proof}

With the target in $|-\rangle=(|0\rangle-|1\rangle)/\sqrt2$, phase kickback
gives the clean phase oracle
$O^{\mathrm{ord}}_{k,J}|s\rangle=(-1)^{g_{k,J}(s)}|s\rangle$, and analogously
for $h_{k,J}$; the target remains in $|-\rangle$ and can be reused. These
oracles use only exact reversible Boolean gates.

\section{Quantum support search and resource analysis}
\label{sec:search}

\subsection{Layerwise minimum finding}
\label{subsec:layerwise}

For $k\ge2$, let $A_k$ prepare the Dicke state
$A_k|0^n\rangle=|D^n_k\rangle=N_k^{-1/2}\sum_{|s|=k}|s\rangle$, using the
deterministic construction of \citet{Bartschi2019}
with $O(nk)$ gates in a model with exact single-qubit rotations. The
reflection
\[
D_k=2|D^n_k\rangle\langle D^n_k|-I=A_k\bigl(2|0^n\rangle\langle0^n|-I\bigr)A_k^\dagger
\]
is the standard amplitude-amplification construction \citep{Brassard2000}; the
central reflection costs $O(n)$ gates. For a temporary incumbent $J$, a
Grover iteration is $D_kO^{\mathrm{ord}}_{k,J}$ and costs $O(nk+\Gamma_k)$
gates, including both $A_k$ and $A_k^\dagger$. The state $|D^n_k\rangle$
covers the full layer, so no prior knowledge of $|\mathcal C_k|$ and no
filtered state preparation are required.

\paragraph{Bounded-budget minimum finding.}
We apply the minimum-finding algorithm of \citet{DurrHoyer1996} to the strict total order $\triangleleft_k$ on
$\mathcal S_k$. A run samples a uniform initial incumbent, and repeatedly
uses the randomized search of \citet{Boyer1998}, which does not
require the number of improving elements, to find a candidate preceding the
current incumbent; each measured candidate is verified classically and
replaces the incumbent if it precedes it. We charge one unit for each
Grover iteration and one unit for each measurement-and-verification attempt.
Let $c_{\mathrm{DH}}$ be a constant such that, under this charging, the
expected number of charged units before the incumbent becomes the least
element is at most $c_{\mathrm{DH}}\sqrt{N_k}$; such a constant follows from
the analysis in \citet{Boyer1998} and \citet{DurrHoyer1996}; an empirical estimate is
reported in Section~\ref{subsec:exp-search}. A run with budget
\[
\mathcal B_k=\bigl\lceil 2c_{\mathrm{DH}}\sqrt{N_k}\,\bigr\rceil
\]
is stopped when the budget is exhausted and returns its current incumbent;
by Markov's inequality it returns the least element of $\triangleleft_k$ with
probability at least $1/2$. An unsuccessful attempt is not treated as a
certificate of optimality, and measured strings outside $\mathcal S_k$ (which
have probability zero in the ideal algorithm) are treated as unsuccessful
attempts.

\paragraph{Algorithm.}
For $U\ge2$ and $\delta\in(0,1/2]$, let
$R=\lceil\log_2((U-1)/\delta)\rceil$.
\begin{enumerate}[leftmargin=*]
\item Initialize the global incumbent $J$ as a minimum diagonal entry of
$\Qh$, breaking ties by $\nu$.
\item For $k=2,\dots,U$, perform $R$ independent bounded-budget runs over
$\mathcal S_k$ and let $\widehat I_k$ be the best returned support under
$\triangleleft_k$.
\item If $\widehat I_k$ is admissible and $\widehat I_k\prec J$, replace $J$
by $\widehat I_k$.
\item Return $J$ and the rational point of
Proposition~\ref{prop:integer-admissibility}.
\end{enumerate}
For $U=1$, the diagonal scan gives a global minimizer. If
$\mathcal C_k=\emptyset$, every returned support is rejected by exact
verification and the global incumbent is unchanged; the algorithm need not
certify that a layer is empty.

\subsection{Correctness and explicit resource bounds}
\label{subsec:explicit-search-bounds}

In this subsection, elementary gates are NOT, CNOT, Toffoli and arbitrary
single-qubit rotations, the latter treated as exact.

\begin{theorem}[Explicit quantum support search for rational StQP]\label{thm:explicit-quantum-stqp}
Let $Q\in\mathbb Q^{n\times n}$ be symmetric, $\Qh=DQ$, and let
$U\in\{1,\dots,n\}$ be a valid bound on $\omega(G)$. For every
$\delta\in(0,1/2]$, the algorithm returns a rational global minimizer of StQP
with probability at least $1-\delta$, without constructing the family of
admissible supports. For $U\ge2$, its quantum gate count and number of qubits
are
\[
G_{\mathrm{search}}=O\Bigl(R\sum_{k=2}^U\sqrt{\tbinom nk}\,(nk+\Gamma_k)\Bigr),
\qquad
n_{\mathrm{qubits}}=O\Bigl(n+\max_{2\le k\le U}\Xi_k\Bigr),
\]
and the number of classical candidate verifications in layer $k$ is
$O\bigl(R\sqrt{\binom nk}\bigr)$.
\end{theorem}

\begin{proof}
By Proposition~\ref{prop:integer-scaling}, scaling preserves admissible
supports, minimizers and orderings. By
Theorem~\ref{thm:explicit-comparison-oracles}, the phase oracle implements
$\triangleleft_k$ exactly on $\mathcal S_k$, so each run returns the least
element with probability at least $1/2$. With $R$ independent runs, a
nonempty admissible layer fails with probability at most
$2^{-R}\le\delta/(U-1)$; a union bound over the $U-1$ layers gives total
failure probability at most $\delta$. If no layer fails, comparison with the
singleton incumbent yields a support attaining $v^*$ by \eqref{eq:vstar}, and
Proposition~\ref{prop:integer-admissibility} gives the rational minimizer.
Each run uses at most $\mathcal B_k=O(\sqrt{N_k})$ charged units, each costing
$O(nk+\Gamma_k)$ gates, including initial attempt preparations. Layers and
runs are sequential, so workspaces are reused and the qubit count is a
maximum over layers.
\end{proof}

\begin{corollary}[Cardinality-constrained StQP]
\label{cor:cardinality-search}
Under the assumptions of Theorem~\ref{thm:explicit-quantum-stqp},
let $K\in\{1,\dots,n\}$ and set $M=\min\{K,U\}$.
For every $\delta\in(0,1/2]$, the layerwise algorithm,
restricted to layers $k\le M$, returns an exact rational
global minimizer of \eqref{eq:stqpK} with probability
at least $1-\delta$.

For $M\ge2$, let
\[
R_M=\left\lceil
\log_2\frac{M-1}{\delta}
\right\rceil,
\]
and let $\Gamma_k^{(M)}$ and $\Xi_k^{(M)}$ denote
the quantities $\Gamma_k$ and $\Xi_k$ with $W_U$
replaced by $W_M$.
The quantum gate count and number of qubits satisfy
\[
G_{\mathrm{search},K}
=
O\!\left(
R_M\sum_{k=2}^{M}
\sqrt{\binom{n}{k}}\,
\bigl(nk+\Gamma_k^{(M)}\bigr)
\right),
\qquad
n_{\mathrm{qubits},K}
=
O\!\left(
n+\max_{2\le k\le M}\Xi_k^{(M)}
\right).
\]
The number of classical candidate verifications in
layer $k$ is
$O\!\left(R_M\sqrt{\binom{n}{k}}\right)$.
For $M=1$, a classical diagonal scan returns a global
minimizer, and no quantum search is required.
\end{corollary}

\begin{proof}
The support characterization of \eqref{eq:stqpK}
ensures that a global minimizer is attained on an
admissible support of cardinality at most $K$.
Every admissible support is a clique of the convexity
graph and therefore has cardinality at most $U$.
Consequently, it suffices to search layers
$k\le M=\min\{K,U\}$.
The proof of Theorem~\ref{thm:explicit-quantum-stqp}
then applies with $U$ replaced by $M$:
all candidate and incumbent supports have cardinality
at most $M$, so width $W_M$ suffices for the scalar
support data and comparisons.
The stated bounds follow, with a union bound over
the $M-1$ searched layers when $M\ge2$.
If $M=1$, an optimal singleton is obtained by
minimizing the diagonal entries.
\end{proof}

For $K\ge2$ constant independently of $n$, the number of charged candidate
queries is $O(n^{K/2}\log(1/\delta))$, and the gate and qubit bounds are
polynomial in $n$ and in the input bit length. This applies, in particular,
to cardinality-constrained minimum-variance portfolio selection with a fixed
maximum number of assets. The comparison with classical methods is
discussed in Section~\ref{subsec:comparison-baseline}.

\paragraph{Classical processing.}
For each measured candidate, the controller verifies admissibility and, when
admissible, computes $(p_I,\delta_I)$ and the incumbent comparison by the
exact procedures above, with $O(\Gamma_k)$ bit operations; hence
\[
T_{\mathrm{candidates}}=O\Bigl(R\sum_{k=2}^U\sqrt{\tbinom nk}\,\Gamma_k\Bigr).
\]
For the winning support $I$, $x_{i_\ell}=z_{I,\ell}/\delta_I$ and
$v^*_Q=p_I/(D\delta_I)$. The one-time costs $T_{\mathrm{setup}}$ of
constructing $D$, $\Qh$, the convexity graph and the chosen $U$, and of
generating the lookup, arithmetic and state-preparation circuit descriptions,
are reported separately. The choice $U=n$ requires no graph optimization;
the cost of computing a tighter bound is part of $T_{\mathrm{setup}}$.
Generating a circuit description is not equivalent to executing its gates,
and no wall-clock speedup is inferred by adding these different measures.

\noindent For the cardinality-constrained variant of
Corollary~\ref{cor:cardinality-search}, the same
candidate-processing bound holds with $U$, $R$ and
$\Gamma_k$ replaced by $M$, $R_M$ and
$\Gamma_k^{(M)}$, respectively; setup costs remain
accounted for separately.

\subsection{Finite-gate implementation}
\label{subsec:finite-gate-synthesis}

The arithmetic, lookup and comparison circuits are exact over
$\{\mathrm{NOT},\mathrm{CNOT},\mathrm{Toffoli}\}$, which admits exact
Clifford+$T$ decompositions \citep{NielsenChuang2010}. Only the rotations in
the Dicke-state preparations must be approximated. Run the ideal algorithm
with failure parameter $\delta/2$, i.e.\ with
$R_*=\lceil\log_2(2(U-1)/\delta)\rceil$. Let $\rho_k=O(nk)$ be the number of
elementary rotations in $A_k$ (after decomposing controlled rotations). Since
a run performs at most $\mathcal B_k$ Grover iterations, each with $A_k$ and
$A_k^\dagger$, and at most $\mathcal B_k$ initial preparations,
\[
L_{\mathrm{rot}}=\max\Bigl\{1,\ \sum_{k=2}^UR_*\rho_k(3\mathcal B_k+1)\Bigr\}
=O\Bigl(1+R_*\sum_{k=2}^Unk\sqrt{\tbinom nk}\Bigr)
\]
bounds all rotation occurrences in the worst case. Each rotation is
approximated to operator-norm accuracy
$\varepsilon_{\mathrm{rot}}=\delta/(4L_{\mathrm{rot}})$, and the inverse
preparation uses the reversed and inverted synthesized sequences.

\begin{theorem}[Finite-gate implementation]\label{thm:finite-gate-stqp}
Let $s(\varepsilon)$ bound the length of a finite-gate approximation of one
single-qubit rotation to accuracy $\varepsilon$. Under the assumptions of
Theorem~\ref{thm:explicit-quantum-stqp}, the synthesized algorithm returns a
rational global minimizer with probability at least $1-\delta$, with gate count
\[
G_{\mathrm{finite}}=O\Bigl(G^*_{\mathrm{ideal}}+L_{\mathrm{rot}}\,
s\bigl(\delta/(4L_{\mathrm{rot}})\bigr)\Bigr),
\]
where $G^*_{\mathrm{ideal}}$ is the bound of
Theorem~\ref{thm:explicit-quantum-stqp} with $R$ replaced by $R_*$. With
ancilla-free synthesis the qubit bound is unchanged. Using a constructive
Solovay--Kitaev procedure \citep{DawsonNielsen2006},
$s(\varepsilon)=O(\log^c(1/\varepsilon))$ for a constant $c$, so
\[
G_{\mathrm{finite}}=O\Bigl(G^*_{\mathrm{ideal}}+L_{\mathrm{rot}}\log^c\tfrac{4L_{\mathrm{rot}}}{\delta}\Bigr).
\]
\end{theorem}

\begin{proof}
The ideal algorithm fails with probability at most $\delta/2$. By
Lemma~\ref{lem:accumulated-synthesis-error} in Appendix~\ref{app:synthesis},
which applies a hybrid argument \citep{BernsteinVazirani1997} in diamond norm
\citep{Kitaev2002} and covers intermediate measurements and classical
feed-forward, the output distributions of the ideal and synthesized
algorithms differ in total variation by at most
$2L_{\mathrm{rot}}\varepsilon_{\mathrm{rot}}=\delta/2$. The gate count follows
by replacing each rotation by its synthesized sequence.
\end{proof}

Since the required precision has $O(\log L_{\mathrm{rot}}+\log(1/\delta))$ bits,
it is polynomial in the input parameters even when $L_{\mathrm{rot}}$ is
exponential. Correctly identified supports yield exact rational coordinates;
synthesis error affects only the success probability. The result concerns
ideal logical gates and excludes physical noise and error correction.

\subsection{Comparison baseline}
\label{subsec:comparison-baseline}

The natural enumeration baseline for the support representation examines
$N_{\mathrm{cand}}=\sum_{k=2}^U\binom nk$ candidate supports in addition to the
singleton layer; this is the size of the full candidate domain, not the number
of admissible supports. Within each layer, minimum finding reduces the
number of charged candidate queries from $N_k$ to $O(\sqrt{N_k})$, and the
total is
\[
O\Bigl(R\sum_{k=2}^U\sqrt{\tbinom nk}\Bigr),\qquad
\sum_{k=2}^U\sqrt{\tbinom nk}\le\sqrt{(U-1)N_{\mathrm{cand}}}
\]
by the Cauchy--Schwarz inequality. For fixed $U\ge2$, $N_{\mathrm{cand}}=O(n^U)$
and the charged-query count is $O(n^{U/2}\log(1/\delta))$. When $U=n$,
$N_{\mathrm{cand}}=2^n-n-1$ and the sum contains the central term
$\sqrt{\binom n{\lfloor n/2\rfloor}}=\Theta(2^{n/2}n^{-1/4})$, so the search
remains exponential in the unrestricted case.

This comparison concerns exhaustive enumeration of the same candidate domain. In the non-convex case, classical approaches typically rely on branch-and-bound methods \citep{Liuzzi2019,ScozzariTardella2008} or local-search procedures \citep{Bisori2022}, and their effectiveness is generally assessed empirically through running times on benchmark instances of increasing size. Local-search methods may converge to locally optimal solutions without providing certificates of global optimality, while branch-and-bound algorithms can require prohibitive computational effort on challenging instances before termination. By contrast, for cardinality-constrained formulations, whenever a cardinality bound $K$ is available, either from the underlying application or from a preliminary heuristic, the search procedure of Corollary \ref{cor:cardinality-search} admits an explicit resource bound that depends only on $K$ and is independent of both the instance structure and any pruning mechanism. When $K$ is obtained heuristically, the resulting optimality guarantee applies only to problem \eqref{eq:stqpK}, and not necessarily to the unrestricted problem. Thus, the main contribution is not a practical computational speedup, but rather a fully explicit, resource-accounted benchmark (in particular in the non-convex case) against which future quantum implementations may be evaluated.

\section{Computational validation}
\label{sec:experiments}

The experiments test the integer predicates against an independent rational implementation, verify generated reversible circuits, and quantify their resources. They do not compare runtimes with classical StQP solvers. Implementation details, instance families, and supplementary results appear in Appendix \ref{app:experiments}. A computational replication package containing the code, instance generators, reference results, and reproduction instructions is available upon reasonable request.

\subsection{Implementation and predicate validation}
\label{subsec:exp-setup}
\label{subsec:exp-predicates}
The experiments were run on a Dell XPS 15 equipped with an Intel Core i9-11900H processor
using a 64-bit installation of Python 3.11.0. All predicate calculations use exact integer arithmetic.
The independent reference uses rational symmetric elimination to test
positive definiteness and Gauss--Jordan elimination to recover allocations,
without determinant formulas or integer scaling. Eleven matrix families
include integer and rational random instances, low-rank semidefinite
matrices, circulant matrices, and instances designed to produce singularity,
boundary coordinates, or ties. Motzkin--Straus instances provide the external
check $v^*=1/\omega(G)$ \citep{MotzkinStraus1965}.

Across 660 instances with $n=3,\dots,8$, all 54{,}780 nonempty supports
agreed with the rational reference. We also checked the integer-size bounds,
89{,}740 admissible-support comparisons, and 129{,}360 layer-order comparisons.
The tests include 2{,}387 supports that would be incorrectly accepted if
positive definiteness were tested using only $\delta_I>0$, and 354 admissible
supports with singular $Q_I$. These cases support the need for all leading
minors and for the reduced system rather than inversion of $Q_I$.
The breakdown by family is in Appendix Table~\ref{tab:predicates}.

\subsection{Circuit verification and resources}
\label{subsec:exp-oracles}
\label{subsec:exp-resources}
For 17 instances with $n\in\{4,5,6\}$ and $U=\min\{n,4\}$, we generated
102 comparison-oracle circuits over NOT, CNOT, and Toffoli gates. Exact
bit-sliced simulation checked every incidence string and permitted incumbent,
including inadmissible incumbents and strings outside the layer. In all
50{,}128 tested pairs, the target agreed with the predicate, input and
incumbent registers were preserved, and all workspace returned to zero.
For the tested instances, the resulting basis-state action extends to
superpositions by linearity. Across the 102 verification circuits, the maximum Toffoli count
is $1{,}049{,}792$ and the maximum qubit count is $523{,}286$.

Table~\ref{tab:resources} reports the unoptimized resource sweeps, with
one history qubit per Boolean gate and sequential workspace reuse.
For $k\geq4$ in the cardinality sweep, the evaluator accounts for approximately $95\%$ or more of the Toffoli count. Its count normalized by the exact number of fraction-free updates and squared register width is between 19.7 and 30.0. Increasing $B$ produces resource counts that grow approximately quadratically with the register width; increasing $n$ raises the loader share from $0.8\%$ at $n=4$ to $23\%$ at $n=32$, with the evaluator unchanged. These results identify arithmetic and coherent loading as distinct targets for improvement. Appendix~\ref{sup:subsec:exp-resources} gives the
normalization formula and further analysis.

\begin{table}[htbp]
\centering
\small
\setlength{\tabcolsep}{4pt}
\begin{tabular}{rrrrrrrrrrr}
\toprule
$n$ & $k$ & $B$ & $W_U$ & Toffoli & CNOT & Qubits & Loader & Evaluator & Comparator & $\kappa_k$\\
\midrule
8 & 2 & 3 & 18 & 5{,}310 & 13{,}755 & 1{,}942 & 11\% & 56\% & 32\% & --\\
8 & 3 & 3 & 31 & 46{,}566 & 128{,}241 & 16{,}361 & 2\% & 86\% & 12\% & 29.5\\
8 & 4 & 3 & 45 & 295{,}106 & 823{,}507 & 148{,}313 & 0.6\% & 95\% & 4\% & 20.9\\
8 & 5 & 3 & 58 & 1{,}135{,}634 & 3{,}134{,}413 & 621{,}168 & 0.2\% & 98\% & 1.7\% & 19.8\\
8 & 6 & 3 & 72 & 3{,}578{,}466 & 9{,}788{,}043 & 2{,}023{,}512 & 0.1\% & 99\% & 0.9\% & 19.7\\
4 & 3 & 3 & 31 & 45{,}790 & 124{,}681 & 16{,}353 & 0.8\% & 87\% & 12\% & 29.5\\
16 & 3 & 3 & 31 & 49{,}590 & 144{,}597 & 16{,}377 & 8\% & 80\% & 11\% & 29.5\\
32 & 3 & 3 & 31 & 59{,}350 & 207{,}917 & 16{,}409 & 23\% & 67\% & 9\% & 29.5\\
8 & 3 & 8 & 56 & 145{,}858 & 402{,}499 & 50{,}732 & 0.9\% & 87\% & 13\% & 29.8\\
8 & 3 & 16 & 96 & 420{,}622 & 1{,}154{,}083 & 145{,}141 & 0.3\% & 87\% & 13\% & 30.0\\
\bottomrule
\end{tabular}
\caption{Generated resource counts of the clean oracle for $g_{k,J}$ (gate
counts include forward and inverse parts). Loader, Evaluator and Comparator
give the shares of the Toffoli count. $\kappa_k$ is the evaluator Toffoli count divided by
$\mathcal U'_k\,w'_{k}(k,B)^2$, where
$\mathcal U'_k=(k-2)(k-1)k/3$ is the exact number of elimination updates of
Proposition~\ref{prop:single-pass}, and
$w'_{k}(k,B)=2t_k+\lceil\log_2(k+1)\rceil+2$ is the register width of
Proposition~\ref{prop:single-pass}(d).}
\label{tab:resources}
\end{table}

\subsection{Minimum-finding simulations}
\label{subsec:exp-search}
A state-vector check with $n=6$, $k=3$, and all 20 incumbents reproduced the
Grover measurement law to within $3\cdot10^{-15}$. Subsequent search
experiments used this law rather than full state-vector simulation of the
large generated circuits. Across 4{,}000 runs per domain size, the mean
charged cost approached $2.9\sqrt N$. This is an empirical estimate, not a
certified universal bound for the constant in Section~\ref{subsec:layerwise}.
Using this estimate, complete-algorithm simulations returned a global
minimizer in all 12{,}800 trials on 16 instances. Charged costs exceeded
enumeration for $n\le10$ and fell below it only at $n=12$ in this sample.
The detailed budgets and results are in Appendix
Tables~\ref{tab:dh}--\ref{tab:e2e}.

\section{Conclusions}
\label{sec:conclusions}

Building on known structural results, this work develops a quantum
support-search framework for standard quadratic programming.
The underlying reformulation turns a continuous, possibly nonconvex
optimization problem into a finite search over admissible supports,
while preserving the continuous optimal value. Each admissible
support determines a unique relative-interior face minimizer, and
at least one global minimizer can be recovered from such a support.
Unlike formulations based on discretizing the decision variables,
this representation retains continuous allocations, which are
recovered exactly as rational vectors for rational input.
The convexity graph further links the continuous geometry of the
problem to the combinatorial structure of the search domain by
restricting admissible supports to cliques.

The main constructive contribution is to make this support
representation operational within a quantum algorithm.
Rather than assuming access to coherent admissibility and
objective-comparison predicates, we provide explicit clean
reversible circuits implementing them, including coherent loading
of the instance data, exact fraction-free arithmetic, integer sign
tests, and tie-breaking comparisons.
A fixed-schedule single-pass evaluator computes the quantities
needed by these predicates without row exchanges, using an activity
flag to handle failure and mask the outputs.
Combined with fixed-cardinality state preparation and layerwise
quantum minimum finding, these circuits yield a complete algorithmic
construction that returns an exact rational global minimizer with
prescribed success probability.
The resource analysis accounts for gate counts, workspace,
classical processing, and finite-gate synthesis error.

\noindent A particularly relevant regime arises when the problem is non-convex, as is the case for the cardinality-constrained StQP. Consider, for example, a cardinality bound $K \ge 2$ that remains fixed as the problem dimension grows, as in minimum-variance portfolio selection with a prescribed maximum number of assets. In this setting, both exhaustive support enumeration and our quantum construction have polynomial complexity in the problem dimension and input bit length. Relative to exhaustive enumeration, the quantum search reduces the number of candidate supports that must be examined from $O(n^K)$ to $O(n^{K/2}\log(1/\delta))$, while preserving continuous asset allocations and enabling the exact recovery of rational portfolio weights. Although classical methods may perform efficiently on many instances, their performance typically depends on the specific structure of the problem and they generally do not provide instance-independent complexity guarantees. By contrast, whenever a cardinality bound is available, our approach yields an explicit resource bound that depends only on $K$. Whether this theoretical advantage can translate into practical computational gains remains contingent on future advances in quantum hardware and on further reductions in oracle-construction overhead.

Computational validation confirms agreement between the integer
predicates and an independent rational implementation on the tested
instances. Additional checks validate the single-pass evaluator's
admissibility flag and failure criterion, the agreement of its
certificate quantities with determinant-based certificates on
admissible supports, and its intermediate-value bounds and
elimination-update count on successful executions.
Exhaustive basis-state simulation of the generated verification
circuits also confirms their correctness and cleanliness on the
tested instances.
The resulting resource requirements reveal the considerable overhead imposed by the current conservative construction.

Promising improvements include optimized fraction-free arithmetic,
space-efficient reversible schedules, and more efficient coherent
loading. Quantum linear-system approaches \citep{Harrow2009,Morales2026} would need
to be assessed against this exact baseline, including their precision costs.
A central question for future work is whether the convexity graph
can be exploited not only to bound support cardinalities, but also
to restrict quantum search to smaller, efficiently accessible
candidate domains, while accounting for the cost of constructing
and accessing those domains.
Establishing a computational advantage will require comparing the
resulting implementations with classical algorithms exploiting
the same structure.


\appendix

\section{Proofs of Section~\ref{sec:integer} and Section~\ref{sec:reversible}}
\label{app:proofs}

\subsection{Proof of Proposition~\ref{prop:certificate-bit-length}}
\label{app:bitlength}

Every entry of $\Sh_I$ is a sum of four signed entries of $\Qh$, so
$|(\Sh_I)_{ab}|\le4H$; similarly $|(\ch_I)_a|\le2H$ and $|\dhat_I|\le H$. By
Hadamard's inequality \citep{HornJohnson2013}, a $j\times j$ matrix with entries
bounded by $4H$ has determinant at most $j^{j/2}(4H)^j\le r^{r/2}(4H)^r\le T_k$
for $j\le r$. This bounds $\delta_{I,j}$ and, since replacement matrices have
entries bounded by $4H$, also $z_{I,\ell}$ for $\ell\le r$. Then
$|z_{I,k}|\le T_k+rT_k=kT_k$ and
$|p_I|\le HT_k+2rHT_k=(2k-1)HT_k$. Since $T_k$ is nondecreasing in $k$ and
the singleton conventions satisfy the same bounds, for $|I|,|J|\le U$ we have
$|F(I,J)|\le|p_I||\delta_J|+|p_J||\delta_I|\le2(2U-1)HT_U^2=R_U$; partial sums
and individual products are bounded by the same quantity. A signed
two's-complement register of width $W_U$ represents $[-R_U,R_U]$, and
$\log_2R_U=1+\log_2(2U-1)+B+2t_U$ gives the asymptotic bound.
\hfill$\square$

\subsection{Proof of Proposition~\ref{prop:single-pass}}
\label{app:single-pass}
\begin{proof}
Let $D_j=\delta_{I,j}$, $D_0=1$. \emph{Invariant.} At the beginning of stage
$s$, if $D_1,\dots,D_{s-1}>0$ and $d=D_{s-1}$, then for all $i,j\ge s$
($j\le r+1$)
\[
A_{ij}=\det N\bigl[(1,\dots,s-1,i),(1,\dots,s-1,j)\bigr].
\]
It holds for $s=1$. If $A_{ss}=D_s>0$, Sylvester's determinant identity
\citep{Bareiss1968}, which holds for every matrix and every choice of the
extra row and column, gives
$A_{ss}A_{ij}-A_{is}A_{sj}=D_{s-1}\det N[(1,\dots,s,i),(1,\dots,s,j)]$.
Hence each division is exact, and the invariant holds at stage $s+1$ with
$d=D_s$. In particular $A_{ss}=D_s$ at stage $s$ as long as no FAIL occurred
before. This proves (b), once (a) is shown.

(a) By Sylvester's criterion, $S\succ0$ iff $D_j>0$ for all $j$. If
$S\succ0$, every pivot test passes by the invariant. If not, let $j$ be the
smallest index with $D_j\le0$; stages $1,\dots,j-1$ run as above, and at stage
$j$ the pivot equals $D_j\le0$, so the procedure returns FAIL.

(c) Each update replaces row $i$ by $(A_{ss}/d)$ times row $i$ minus a
multiple of row $s$, with $A_{ss}/d\ne0$, and rows are frozen after their
stage. Hence $[U\,|\,u]=L[S\,|\,c]$ for an invertible rational matrix $L$,
so $y^*=S^{-1}c$ satisfies $Uy^*=u$ with $U$ upper triangular and
$U_{ss}=D_s$. By Cramer's rule, $\delta_r y^*_j=z_{I,j}\in\mathbb Z$. Row $r$
gives $\delta_r y^*_r=u_r$, so $\zeta_r=z_{I,r}$. For $s<r$, multiplying
row $s$ by $\delta_r$ gives
$\delta_s\,(\delta_r y^*_s)=\delta_r u_s-\sum_{j>s}U_{sj}(\delta_r y^*_j)$.
By downward induction $\zeta_j=z_{I,j}$ for $j>s$, hence the right-hand side
equals $\delta_s z_{I,s}$, the division by $\delta_s>0$ is exact and
$\zeta_s=z_{I,s}$. Then $z_{I,k}=\delta_I-\sum_\ell z_{I,\ell}$ and $p_I$ are
obtained as in Section~\ref{subsec:objective}.

(d) Every $A_{ij}$ and every $U_{sj}$ is a minor of $N$ of order at most $r$
with entries bounded by $2^\beta$, so by Hadamard's inequality
\citep{HornJohnson2013} it is at most $r^{r/2}2^{\beta r}\le T_k$. Products
are at most $T_k^2$ and elimination numerators at most $2T_k^2$. In the back
substitution, using $|z_{I,j}|\le T_k$,
$|D_r u_s-\sum_{j>s}U_{sj}\zeta_j|\le T_k^2+(r-s)T_k^2\le rT_k^2$.
Since $2kT_k^2+1\le(2k+1)T_k^2$, a signed register of width
$2t_k+\lceil\log_2(2k+1)\rceil+1\le w'_k$ represents all of them.

(e) On an execution without \textnormal{FAIL}, the number of
elimination updates is
\[
\sum_{s=1}^{r-1}(r-s)(r+1-s)
=
\frac{(r-1)r(r+1)}{3}.
\]
Early termination can only reduce this number.
The back substitution, when reached, uses $r-1$ divisions
and $O(r^2)$ multiplications.
\end{proof}

\section{General pivoted fraction-free determinant evaluation}
\label{app:bareiss-general}

This appendix states the general fraction-free determinant procedure based
on Bareiss' integer-preserving algorithm \citep{Bareiss1968}, with
deterministic row pivoting and an explicit treatment of singular inputs.
Pivoting is required for arbitrary integer matrices, such as the replacement
matrices $\Sh_I^{(\ell)}$, which are not symmetric and may have zero
diagonal pivots. The procedure is not used by the search oracles, which rely
on the single-pass evaluator of Proposition~\ref{prop:single-pass}; it
provides the determinant-based reference used in
Appendix~\ref{app:experiments} and the cost baseline quoted in
Section~\ref{subsec:single-pass-eval}.

\subsection{Procedure}

For $M\in\mathbb Z^{m\times m}$ with $m\ge2$:

\begin{enumerate}[leftmargin=*]
\item Initialize $A\leftarrow M$, $d\leftarrow1$, $\sigma\leftarrow1$.
\item For $s=1,\dots,m-1$:
 \begin{enumerate}
 \item find the smallest $v\in\{s,\dots,m\}$ with $A_{vs}\ne0$; if none
 exists, return $0$;
 \item if $v\ne s$, exchange rows $v$ and $s$ and set $\sigma\leftarrow-\sigma$;
 \item with $p=A_{ss}$, replace, for all $i,j>s$,
 $A_{ij}\leftarrow(pA_{ij}-A_{is}A_{sj})/d$, keeping $p$, $d$, the pivot row
 and the entries $A_{is}$ unchanged during these updates;
 \item set $A_{is}\leftarrow0$ for $i>s$ and then $d\leftarrow p$.
 \end{enumerate}
\item Return $\sigma A_{mm}$.
\end{enumerate}
For $m=1$ return $M_{11}$. The procedure is fraction-free but not
division-free.

The following proposition collects standard properties of Bareiss elimination \citep{Bareiss1968}, together with explicit intermediate-value and bit-complexity bounds used in our resource analysis.
\begin{proposition}[Bareiss elimination: correctness and explicit bounds]\label{prop:bareiss}
For every $M\in\mathbb Z^{m\times m}$, the procedure returns $\det M$, and
every executed division has nonzero divisor and zero remainder. If
$|M_{ij}|\le2^\beta$ and
\[
\tau(m,\beta)=m\beta+\Bigl\lceil\tfrac m2\log_2m\Bigr\rceil,\qquad
V=2^{\tau(m,\beta)},
\]
then every stored entry, pivot and divisor is bounded in absolute value by
$V$, and every numerator before division by $2V^2$; signed registers of width
$w_{\det}(m,\beta)=2\tau(m,\beta)+3$ suffice. The procedure uses $O(m^3)$
arithmetic operations and, with schoolbook multiplication and binary long
division \citep{BrentZimmermann2010},
$O\bigl(m^5(\beta+\log_2(m+1))^2\bigr)$ bit operations.
\end{proposition}

\subsection{Proof of Proposition~\ref{prop:bareiss}}
\label{app:bareiss}

The correctness argument uses the standard bordered-minor
invariant of Bareiss elimination \citep{Bareiss1968};
we recall it briefly to justify the pivoting and singular
branches, and then derive the stated resource bounds.

At the beginning of stage $s$, let $\widetilde M=PM$,
where $P$ is the accumulated row permutation, and let
$D_j=\det\widetilde M[1:j,1:j]$, with $D_0=1$.
The active entries satisfy
\[
A_{ij}
=
\det\widetilde M[
(1,\dots,s-1,i),(1,\dots,s-1,j)],
\qquad
d=D_{s-1}\ne0.
\]
Exchanging active rows preserves this invariant.
For a nonzero pivot $p=A_{ss}=D_s$, Sylvester's
determinant identity shows that each update numerator
is $D_{s-1}$ times the corresponding bordered minor.
Thus every executed division is exact, and the invariant
propagates to the next stage.
If no nonzero pivot exists in the active column, the
Schur complement of the invertible leading block has
a zero column, so $M$ is singular.
Otherwise the final entry is $\det(PM)$, and the
accumulated permutation sign gives $\det M$.
For $s=1$, the same singularity argument applies
directly to the first column of $M$.

Every stored matrix entry, pivot and divisor is a minor
of a row-permuted input matrix, or the initial value $1$.
Hadamard's inequality therefore bounds its absolute
value by
\[
m^{m/2}2^{m\beta}
\le 2^{\tau(m,\beta)}=V.
\]
Each numerator before division has absolute value at
most $2V^2$, so signed registers of width
$2\tau(m,\beta)+3$ suffice.

There are $O(m^3)$ elimination updates, while pivot
selection and row exchanges do not increase this
arithmetic-operation bound.
Since the register width is
$O(m(\beta+\log_2(m+1)))$, schoolbook multiplication
and binary long division \citep{BrentZimmermann2010}
give the total bit-operation bound
\[
O\!\left(m^5(\beta+\log_2(m+1))^2\right).
\]
\hfill$\square$

\subsection{Fixed-schedule reversible implementation}

\paragraph{Fixed-width arithmetic and circuit completion.}
Entries are supplied in a fixed signed encoding of $[-2^\beta,2^\beta]$ and
sign-extended to width $w=w_{\det}(m,\beta)$; bit strings outside this
interval are mapped to output zero by an initial validity test. Addition,
subtraction and multiplication are fixed-width Boolean circuits. Signed
division truncates towards zero for a nonzero divisor and outputs zero for a
zero divisor; other unreachable inputs receive a prescribed output. By
Proposition~\ref{prop:bareiss}, every active update on a valid input has a
nonzero divisor, zero remainder and representable quotient, so these
conventions do not affect valid outputs.

\paragraph{Fixed scheduling of pivoting and singular branches.}
All $m-1$ stages are present. At stage $s$, the circuit computes
$n_v=[A_{vs}\ne0]$ and the first-nonzero selectors
$e_v=n_v\land\bigwedge_{u=s}^{v-1}\neg n_u$ by prefix conjunctions, and
$f_s=\bigvee_ve_v$. For each $v>s$, rows $s$ and $v$ are exchanged, and the
sign is updated, under the control of the activity bit and $e_v$. If
elimination is active and $f_s=0$, a singularity flag is set and the activity
bit is cleared; all later stages are masked. The output is zero on singular
branches and $\sigma A_{mm}$ otherwise.

\begin{lemma}[Boolean circuit size]\label{lem:Boolean_circuit_size}
The fixed-schedule pivoted determinant circuit has size $O(m^3w^2)$.
\end{lemma}

\begin{proof}
Nonzero tests, comparisons, and word selection cost $O(w)$.
Schoolbook multiplication and restoring long division cost
$O(w^2)$ \citep{BrentZimmermann2010}.
There are $O(m^3)$ elimination updates, each of cost $O(w^2)$.
Conditional row exchanges with every candidate row contribute
$O(m^3w)$ gates; pivot selection, activity tests, masking, and
output selection do not increase the overall bound.
The total size is therefore $O(m^3w^2)$.
\end{proof}

\section{Reversible implementation details}
\label{app:reversible}

\begin{proof}[Proof of Lemma~\ref{lem:clean-reversible-embedding}]
Write the Boolean circuit over AND, XOR, and NOT gates and
evaluate it in topological order, assigning a fresh
zero-initialized wire to every gate output.
An AND gate is implemented by a Toffoli gate, an XOR gate by
two CNOT gates, and a NOT gate by a CNOT gate and a NOT gate.
The resulting forward circuit $V$ preserves the input $x$
and stores $f(x)$ together with all intermediate values.
Apply $b$ CNOT gates from the output wires to the target
register $y$, thereby replacing $y$ by $y\oplus f(x)$.
Then apply $V^\dagger$, which does not act on $y$ and restores
all intermediate wires to zero \citep{Bennett1973}.
The construction uses $O(S+b)$ gates and $O(S+b)$ clean
work qubits.
\end{proof}

\paragraph{Single-pass evaluator.}
All $r=k-1$ stages are present. At stage $s$, the circuit computes
$\pi_s=\mathrm{Act}_s\wedge[A_{ss}>0]$ and sets
$\mathrm{Act}_{s+1}=\pi_s$, with $\mathrm{Act}_1=1$.
Once cleared, the activity flag remains zero.
There are no row exchanges, no first-nonzero selectors, and no
permutation sign.

In the generated implementation, arithmetic operations are evaluated
on a fixed schedule, including after the activity flag has been
cleared. The total fixed-width arithmetic defines these inactive
computations, and all output minors and Cramer numerators are masked
by the final activity flag. Thus intermediate values after failure
are irrelevant to the output.

On active elimination updates, every divisor is
$\delta_{I,s-1}>0$, with $\delta_{I,0}=1$
(Proposition~\ref{prop:single-pass}). Every such division is exact, and the intermediate
values satisfy the bounds of Proposition~\ref{prop:single-pass}(d).
If the final activity flag is one, the back substitution also uses
positive divisors and exact divisions, and returns the Cramer
numerators. If the final activity flag is zero, all output minors
and Cramer numerators are zero. The subsequent formulas therefore
also give $z_{I,k}=0$, $p_I=0$, and $a_k(I)=0$.

There are $(k-2)(k-1)k/3$ elimination-update slots and
$O(k^2)$ further arithmetic operations.
Using width $w'_k$ for elimination and back substitution gives
Boolean circuit size $O(k^3(w'_k)^2)$ for these operations.
The final support quantities and their output encodings are
handled at width $w_k$, as defined in Section~\ref{subsec:reversible-support-data}.
Since $w'_k\leq w_k$, the complete support-data evaluator has
Boolean circuit size $O(k^3w_k^2)$.
Applying Lemma~\ref{lem:clean-reversible-embedding} gives the gate and workspace bounds
$O(k^5(B+\log_2(k+1))^2)$ of Theorem~\ref{thm:reversible-support-data}.

\paragraph{Ordered support extraction.}
For an incidence string $s$, let $c_j=\sum_{u\le j}s_u$ and
$v_k(s)=[c_n=k]$. For $t=1,\dots,k$, the address $i_t(s)$ is the unique $j$
with $s_j=1$ and $c_j=t$ if $v_k(s)=1$, and $0$ otherwise. It is computed by
XORing the encoding of $j$ into the $t$th address register under the control
of $v_k(s)\land s_j\land[c_j=t]$. The addresses are ordered because $c_j$ is
nondecreasing in $j$.
The prefix counts use clean fixed-width adders on $\ell$-bit registers.
Each selector $v_k(s)\land s_j\land[c_j=t]$ is computed, used to XOR the
encoding of $j$ into the $t$th address register, and uncomputed. For valid
strings exactly one selector contributes to each address, for invalid ones
none. This costs $O(nk\ell^2)$ gates and $O(n\ell^2+k\ell)$ temporary qubits,
including the uncomputation of prefix data.

\section{Synthesis-error accounting}
\label{app:synthesis}

\begin{lemma}[Accumulated synthesis error]\label{lem:accumulated-synthesis-error}
If every rotation occurrence is approximated to operator-norm error at most
$\varepsilon_{\mathrm{rot}}$ and every execution contains at most
$L_{\mathrm{rot}}$ occurrences, the final classical output distributions of
the ideal and synthesized algorithms differ in total variation by at most
$2L_{\mathrm{rot}}\varepsilon_{\mathrm{rot}}$, also in the presence of
intermediate measurements and classical feed-forward.
\end{lemma}

\begin{proof}
If $\|V-\widetilde V\|_{\mathrm{op}}\le\varepsilon$, the induced channels
differ by at most $2\varepsilon$ in diamond norm \citep{Kitaev2002}. Replace
the occurrences one at a time; since all other operations, including
measurements and feed-forward, are channels of diamond norm one, each
replacement changes the complete channel by at most
$2\varepsilon_{\mathrm{rot}}$ \citep{BernsteinVazirani1997}. A bounded adaptive
execution can be padded with identities, so the uniform bound
$L_{\mathrm{rot}}$ applies. The final measurement does not increase
distinguishability.
\end{proof}

The inverse of a synthesized sequence is obtained by reversing and inverting
it, so $\|R^\dagger-\widetilde R^\dagger\|_{\mathrm{op}}=\|R-\widetilde R\|_{\mathrm{op}}$.
The accuracy requirement includes both angle generation and synthesis, for
instance with $\varepsilon_{\mathrm{rot}}/2$ each, and approximations use a
consistent global-phase convention so that controlled operations do not
introduce uncontrolled relative phases. Only distinct rotations need to be
synthesized; their compilation is part of $T_{\mathrm{setup}}$.


\section{Detailed computational validation}\label{app:experiments}
\label{sup:sec:experiments}

The results of Sections~\ref{sec:integer}--\ref{sec:search} are exact
statements, and their correctness does not depend on experiments. The purpose
of this section is different: we verify, on small instances, that a direct
implementation of the constructions behaves as stated, and we report the
resource counts of the generated circuits. In particular, we check that the
integer predicates agree with an independent rational implementation, that
the composed reversible oracles compute the intended predicates and are
clean, and that the generated gate and qubit counts follow the bounds. We
also quantify the constants of the bounded-budget minimum finding of
Section~\ref{subsec:layerwise}. No runtime comparison with classical StQP
solvers is attempted, consistently with Section~\ref{subsec:comparison-baseline}.

\subsection{Implementation and test instances}
\label{sup:subsec:exp-setup}

The experiments were run on a Dell XPS 15 equipped with an Intel Core i9-11900H processor
using a 64-bit installation of Python 3.11.0. The integer predicates and reversible arithmetic circuits use exact integer arithmetic, while the independent reference
uses exact rational arithmetic. Every table below is reproduced by a single script.
\begin{enumerate}[leftmargin=*]
\item \emph{Integer predicates.} We scale the input using $D$ equal to the least common multiple
of the denominators (Section~\ref{subsec:rational-input}), implement the determinant procedure of
Appendix~\ref{app:bareiss-general}, checking that every division has a nonzero divisor and zero remainder, and compute
the quantities $\delta_{I,j}$, $z_{I,\ell}$, $p_I$, $a_k$, $F$, $\beta$, $h_{k,J}$, and $g_{k,J}$ defined in
Section~\ref{sec:integer}.
\item \emph{Rational reference.} We independently implement the definitions of Section~\ref{subsec:admissible} using the original rational
matrix $Q$, without relying on determinants, Sylvester's criterion, Cramer's rule, or integer scaling.
We test positive definiteness of $S_I$ by symmetric Gaussian elimination without pivoting, requiring all pivots to be
strictly positive. When $S_I$ is positive definite, we compute $y_I^*$ by Gauss--Jordan elimination with exact rational arithmetic
and check whether the corresponding vector $u_I^*$ is strictly positive. For admissible supports, we evaluate $w(I)$ using both
$d_I-c_I^T y_I^*$ and $u_I^{*T}Q_Iu_I^*$.
\item \emph{Circuit generator.}
We generate bounded-fan-in Boolean circuits over $\{\mathrm{AND},\mathrm{XOR},\mathrm{NOT}\}$ for the single-pass support-data evaluator of Proposition \ref{prop:single-pass}, the loader of Proposition \ref{prop:clean-support-matrix-loader}, and the comparator of Section~\ref{subsec:reversible-comparison}. The evaluator uses pivot-free fraction-free elimination of the augmented matrix $[\hat{S}_I|\hat{c}_I]$, followed by fraction-free back substitution. Arithmetic operations use schoolbook multiplication and restoring long division, with a total signed-division circuit.
We convert each block into a reversible circuit over $\{\mathrm{NOT},\mathrm{CNOT},\mathrm{Toffoli}\}$
using Lemma~\ref{lem:clean-reversible-embedding}, with one work qubit for each Boolean gate. Once a block has been uncomputed, its work qubits are
available for reuse by subsequent blocks.
To construct the oracle, we first apply $\mathcal P_k$ and the forward evaluator $V_k$. We then use the comparator to compute the predicate
into the target qubit and apply $V_k^\dagger$ followed by $\mathcal P_k^\dagger$ to restore the work registers to zero.
This implementation is equivalent to the copy-and-reverse construction described in Section~\ref{subsec:reversible-comparison}.
\item \emph{Simulation.} Since the circuits are classical reversible circuits, their action on computational basis states is simulated exactly
and bit-sliced, i.e.\ on all $2^n$ incidence strings at once. The circuits can be exported in OpenQASM~2.0; for a small oracle we verified that the
exported file is parsed by Qiskit with identical gate counts.
\end{enumerate}

We use eleven families of symmetric matrices, chosen to exercise the boundary
cases of the predicates: random integer matrices with entries in
$\{-1,0,1\}$, $\{-3,\dots,3\}$ and $\{-50,\dots,50\}$; random rational
matrices with numerators in $\{-5,\dots,5\}$ and denominators in
$\{1,\dots,6\}$; positive and negative semidefinite matrices of rank two,
which produce many singular $S_I$; rank-two plus diagonal matrices;
symmetric circulant matrices, whose cyclic symmetry produces exact ties
between objective values; matrices with empty convexity graph; matrices
constructed so that a face minimizer has an exact zero coordinate; and
Motzkin--Straus matrices $Q=I+A_{\bar G}$, where $A_{\bar G}$ is the adjacency
matrix of the complement of a random graph $G$. For the last family the
convexity graph of $Q$ is $G$ itself and the optimal value is $1/\omega(G)$
\citep{MotzkinStraus1965}, which provides an exact external check.

\subsection{Integer predicates}
\label{sup:subsec:exp-predicates}

For $n=3,\dots,8$ we generated ten instances per family and size, 660
instances in total, and examined every nonempty support, 54{,}780 supports
in total. For each support we compared admissibility with the rational
reference; for admissible supports, we also compared the face
minimizer $u_I^*=\delta_I^{-1}z_I$ and the objective value
$w_Q(I)=p_I/(D\delta_I)$.
We also checked the bounds of Propositions~\ref{prop:certificate-bit-length} and~\ref{prop:bareiss}
(stored entries at most $V$, numerators at most $2V^2$).

For every support, we additionally checked the admissibility flag
returned by the single-pass procedure. On admissible supports,
its leading minors, Cramer numerators, and objective numerator
agreed with the determinant-based certificates. We also checked
its failure criterion and, on successful executions, its
intermediate-value bounds and elimination-update count.

For every pair of admissible supports, 89{,}740 pairs, we compared the sign of $F(I,J)$ with
the sign of $w(I)-w(J)$, and for $n\le6$ we compared $g_{k,J}(I)$ with the
reference order $\triangleleft_k$ for all $I,J\in\mathcal S_k$, 129{,}360
pairs. Finally, the global optimum obtained from the integer predicates was
checked against the exact first-order conditions $(Qx^*)_j\ge v^*$ with
equality on the support, against a floating-point multistart
projected-gradient method (which never found a value below $v^*$ by more
than $3\cdot10^{-14}$), and, for Motzkin--Straus instances, against
$1/\omega(G)$.

\begin{table}[htbp]
\centering
\footnotesize
\setlength{\tabcolsep}{3.2pt}
\begin{tabular}{lrrrrrrrrrrr}
\toprule
Family & Inst. & Supp. & Adm. & $\delta_I{=}0$ & Zero & Only & Zero & Adm., & Ties & Row & Err. \\
 & & & & & minor & $\delta_I$ & coord. & $Q_I$ sing. & & exch. & \\
\midrule
Integer $[-1,1]$   & 60 & 4980 &  509 &  616 & 1322 & 140 & 274 &  33 &  439 & 6229 & 0\\
Integer $[-3,3]$   & 60 & 4980 &  620 &  140 &  482 & 187 & 139 &   7 &  140 & 2937 & 0\\
Integer $[-50,50]$ & 60 & 4980 &  679 &    6 &   44 & 262 &   5 &   0 &   10 &  190 & 0\\
Rational           & 60 & 4980 &  621 &   17 &   41 & 233 &  36 &   2 &   35 &  272 & 0\\
PSD rank 2         & 60 & 4980 & 1056 & 2806 & 2806 &   0 & 575 & 239 & 1449 &  768 & 0\\
NSD rank 2         & 60 & 4980 &  330 & 2823 & 2823 & 187 &   0 &   0 &  168 &  744 & 0\\
Rank 2 + diagonal  & 60 & 4980 & 1277 &  175 &  443 &  23 & 283 &  10 &  279 & 2289 & 0\\
Circulant          & 60 & 4980 &  885 &  568 & 1193 & 789 &  30 &  47 & 2945 & 5424 & 0\\
Empty graph        & 60 & 4980 &  330 &  316 &  880 & 312 &   0 &   0 &  162 & 5379 & 0\\
Zero coordinate    & 60 & 4980 &  753 &   75 &  319 & 248 & 152 &  16 &  211 & 2009 & 0\\
Motzkin--Straus    & 60 & 4980 & 1243 & 2127 & 3205 &   6 &   0 &   0 & 5890 & 10426 & 0\\
\midrule
Total              & 660 & 54780 & 8303 & 9669 & 13558 & 2387 & 1494 & 354 & 11728 & 36667 & 0\\
\bottomrule
\end{tabular}
\caption{Validation of the integer predicates against the rational
reference ($n=3,\dots,8$). Columns: supports with $\delta_I=0$; supports with
a zero leading principal minor; supports that would be misclassified as
admissible if only $\delta_I>0$ were tested instead of all leading minors;
supports with $S_I\succ0$ and a zero coordinate of $u_I^*$; admissible
supports with singular $Q_I$; pairs of admissible supports with equal
objective value; row exchanges performed by the determinant procedure;
disagreements of any kind with the reference.}
\label{tab:predicates}
\end{table}

Table~\ref{tab:predicates} reports the results. No disagreement was found.
The boundary cases that motivate the exact treatment occur frequently:
singular reduced matrices ($\delta_I=0$), zero leading minors requiring row
exchanges in the determinant procedure, face minimizers with an exact zero
coordinate, which must be rejected as non-interior, and ties, which are
broken by $\nu$. The column ``Only $\delta_I$'' gives a simple ablation: in
2{,}387 cases $\delta_I>0$ and all Cramer numerators are positive although
$S_I$ is not positive definite, so testing only the determinant of $S_I$
would accept inadmissible supports; all leading minors are needed, as
stated after Proposition~\ref{prop:integer-admissibility}. The 354 admissible
supports with singular $Q_I$ confirm that the reduced system~\eqref{eq:yI},
rather than $Q_I$ itself, is the appropriate object.

\subsection{Reversible oracles}
\label{sup:subsec:exp-oracles}

For 17 instances with $n\in\{4,5,6\}$ drawn from seven of the families above,
and with $U=\min\{n,4\}$, we generated the oracles of
Theorem~\ref{thm:explicit-comparison-oracles} for $h_{k,J}$ and $g_{k,J}$
and every layer $k=2,\dots,U$, 102 circuits in total. Across these circuits, the maximum
Toffoli count is $1{,}049{,}792$ and the maximum qubit count is $523{,}286$. Each circuit was
simulated on all $2^n$ incidence strings for every permitted incumbent: all
$J\in\mathcal S_k$, admissible or not, for $g_{k,J}$, and all admissible $J$
with $|J|\le U$, possibly of a different cardinality, for $h_{k,J}$. Over
the resulting 50{,}128 pairs of incumbent and incidence string, the target bit
always agreed with the integer predicate, including the value zero on strings
of Hamming weight different from $k$, and all other qubits, i.e.\ the
incidence register, the incumbent registers and every work qubit, were
restored to their initial values. This verifies correctness and cleanliness for every incidence string and every permitted incumbent in each tested instance, with workspace initialized to zero. By linearity, the verified action extends to superpositions of these incidence strings.

\subsection{Resource counts}
\label{sup:subsec:exp-resources}

Table~\ref{tab:resources} in Section~\ref{subsec:exp-resources} reports the counts of the generated oracle for
$g_{k,J}$ in three sweeps: in $k$ (with $U=k$), in $n$, and in $B$. The counts
are for the unoptimized construction, i.e.\ one work qubit per Boolean gate
and no measurement-based uncomputation; with the standard decomposition of
the Toffoli gate \citep{NielsenChuang2010}, the $T$-count is at most seven
times the Toffoli count.

For $k\ge4$ the evaluator, and within it the single-pass elimination,
dominates the gate count, in agreement with the term
$k^5(B+\log_2(k+1))^2$ of $\Gamma_k$. The ratio between the Toffoli count and
$\Gamma_k$ is $1.74,4.06,7.62,10.08,12.59$ for $k=2,\dots,6$; this growth is
pre-asymptotic and is explained by the exact operation count. The single-pass elimination
of Proposition~\ref{prop:single-pass} performs
\begin{equation}\label{eq:updates}
\mathcal U'_k=\frac{(k-2)(k-1)k}{3}
\end{equation}
updates for a support of cardinality $k$, i.e.\ $2,8,20,40$ for
$k=3,\dots,6$, a quantity that approaches its asymptotic form $k^3/3$ only
slowly (the ratio is $0.56$ at $k=6$).
Normalized by $\mathcal U'_k$ and by the squared register width,
the evaluator Toffoli count is approximately constant for $k\ge4$ ($19.7$--$20.9$), which is
consistent with the $O(k^3w_k'^2)$ cost of the single-pass evaluator
(Theorem~\ref{thm:reversible-support-data}). In the sweep in $B$, the
counts grow quadratically with the register width, and the ratio to
$\Gamma_k$ increases slightly from 4.07 to 4.21. In the sweep in $n$, the
evaluator count is independent of $n$, while the loader grows from $0.8\%$ to
$23\%$ of the Toffoli count between $n=4$ and $n=32$; most of its cost appears
in CNOT gates, since the table entries are constants and only the address
decoders require Toffoli gates. The qubit counts exceed the classical storage of the elimination by about two
orders of magnitude or more, which reflects the history-preserving embedding
discussed after Theorem~\ref{thm:reversible-support-data}. They follow the
$k^5$ growth of $\Xi_k$ (the ratio to $\Xi_k$ lies between $1.6$ and $7.6$ in
all sweeps). These findings quantify the overhead of the present conservative construction and identify arithmetic evaluation, reversible workspace management, and coherent loading as targets for improvement.


\subsection{Layerwise minimum finding}
\label{sup:subsec:exp-search}

We first checked the measurement law used by the analysis. For an instance
with $n=6$ and $k=3$ ($N_k=20$), and for every temporary incumbent $J$, we
applied the Grover operator $D_kO^{\mathrm{ord}}_{k,J}$ to the Dicke state in
a state-vector simulation, with the phase oracle given by the truth table of
the simulated reversible circuit. Over seven iterations and all 20
incumbents, the probability of measuring a support preceding $J$ agreed with
$\sin^2((2j+1)\theta)$, $\sin^2\theta=t/N_k$, where $t$ is the number of such
supports, to within $3\cdot10^{-15}$.

Since $\triangleleft_k$ is a strict total order on $\mathcal S_k$, the number
of charged units of a minimum-finding run depends only on $N_k$ and not on
the instance. Using this measurement law, we simulated the algorithm of
\citet{DurrHoyer1996} with the randomized search of \citet{Boyer1998} (parameter
$\lambda=6/5$), charging one unit per Grover iteration and per measurement
and verification attempt, including the initial sampling. Table~\ref{tab:dh}
reports the mean number of charged units before the incumbent becomes the
least element, over 4{,}000 runs for each $N$. The normalized mean stabilizes
at about $2.9$, so $c_{\mathrm{DH}}=2.9$ is an empirical estimate of the
constant of Section~\ref{subsec:layerwise} under our charging convention.
With this value, a run with budget $\mathcal B_k=\lceil2c_{\mathrm{DH}}\sqrt{N_k}\rceil$
returned the least element in $95\%$ to $98\%$ of the runs for
$N\in\{16,256,4096,65536\}$. These empirical results do not certify that $2.9$ is a universal bound; the theoretical Markov guarantee requires a valid expectation bound as specified in Section~\ref{subsec:layerwise}.

\begin{table}[htbp]
\centering
\footnotesize
\begin{tabular}{rrrrrrrrrr}
\toprule
$N$ & 2 & 16 & 64 & 256 & 1024 & 4096 & 16384 & 65536\\
\midrule
Mean units$/\sqrt N$ & 1.62 & 2.28 & 2.58 & 2.76 & 2.90 & 2.89 & 2.90 & 2.89\\
95\% quantile$/\sqrt N$ & 4.95 & 5.00 & 5.38 & 5.50 & 5.78 & 5.64 & 5.47 & 5.61\\
\bottomrule
\end{tabular}
\caption{Charged units of a minimum-finding run until the incumbent is the
least element of a totally ordered set of size $N$ (4{,}000 runs per $N$).}
\label{tab:dh}
\end{table}

Finally, we executed the complete algorithm of Section~\ref{subsec:layerwise}
with this measurement law, with $U=\omega(G)$ computed classically, budgets
$\mathcal B_k$ with $c_{\mathrm{DH}}=2.9$, exact classical verification of the
returned supports and the global incumbent update, on 16 instances with
$n\in\{6,8,10,12\}$ from four families, for $\delta\in\{0.25,0.05\}$ and 400
trials per configuration. The returned support was a global minimizer in all
12{,}800 trials. These trials use the empirical constant and do not independently certify the theoretical failure bound.
Table~\ref{tab:e2e} compares, for $\delta=0.05$, the mean number of charged
units with the number $N_{\mathrm{cand}}$ of candidates of exhaustive
enumeration. For $n\le10$ the charged units exceed $N_{\mathrm{cand}}$, owing
to the constant $c_{\mathrm{DH}}$, the budget factor $2$ and the $R$
repetitions; in our sample the ratio falls below one only at $n=12$.
These charged-unit counts do not include the gate-level cost
of coherent oracle evaluation, state preparation, or classical
verification.

\begin{table}[htbp]
\centering
\footnotesize
\begin{tabular}{lrrrrrr}
\toprule
Family & $n$ & $U$ & $R$ & Success & Charged units & $N_{\mathrm{cand}}$\\
\midrule
Rank 2 + diagonal & 6 & 5 & 7 & 400/400 & 228.0 & 56\\
                  & 8 & 5 & 7 & 400/400 & 499.5 & 210\\
                  & 10 & 8 & 8 & 400/400 & 1714.5 & 1002\\
                  & 12 & 10 & 8 & 400/400 & 3966.0 & 4070\\
Motzkin--Straus   & 6 & 3 & 6 & 400/400 & 115.2 & 35\\
                  & 8 & 4 & 6 & 400/400 & 315.5 & 154\\
                  & 10 & 4 & 6 & 400/400 & 508.6 & 375\\
                  & 12 & 4 & 6 & 400/400 & 743.0 & 781\\
Integer $[-3,3]$  & 6 & 2 & 5 & 400/400 & 43.6 & 15\\
                  & 8 & 3 & 6 & 400/400 & 186.1 & 84\\
                  & 10 & 5 & 7 & 400/400 & 910.0 & 627\\
                  & 12 & 5 & 7 & 400/400 & 1436.0 & 1573\\
Rational          & 6 & 3 & 6 & 400/400 & 113.5 & 35\\
                  & 8 & 4 & 6 & 400/400 & 317.9 & 154\\
                  & 10 & 5 & 7 & 400/400 & 905.3 & 627\\
                  & 12 & 6 & 7 & 400/400 & 2018.4 & 2497\\
\bottomrule
\end{tabular}
\caption{Complete layerwise algorithm with $\delta=0.05$: successful trials,
mean charged units over all layers and runs, and number of candidate supports
$N_{\mathrm{cand}}=\sum_{k=2}^U\binom nk$ of exhaustive enumeration.}
\label{tab:e2e}
\end{table}

\end{document}